\documentclass[aip,apl,reprint]{revtex4-2}

\newcommand{\red}[1]{{\color{red}#1}}

\usepackage{xcolor,textcomp}
\usepackage{graphicx,hyperref,epstopdf,mathtools}
\usepackage{amsmath,amssymb,multirow,gensymb,xfrac} 
\usepackage{fancyhdr,float}
\usepackage{array,tabularx,booktabs}
\usepackage[none]{hyphenat}
\usepackage{enumitem,relsize}
\usepackage{footmisc}
\usepackage{soul,subfigure}

\begin{document}

\makeatletter
\let\@fnsymbol\@fnsymbol@latex
\@booleanfalse\altaffilletter@sw
\makeatother

% \preprint{AIP/123-QED}

\title{Pattern Formation in Robotic Mechanical Metamaterial}

\author{Vinod Ramakrishnan}
\affiliation{Department of Mechanical and Aerospace Engineering, University of California, San Diego, La Jolla, California 92093, USA}
\affiliation{Program in Materials Science and Engineering, University of California, San Diego, La Jolla, California 92093, USA}
\affiliation{The Grainger College of Engineering, Department of Mechanical Science and Engineering, University of Illinois, Urbana-Champaign, Urbana, Illinois 61801, USA}

\author{Michael J. Frazier}
\email[Corresponding author email: ]{mjfrazier@ucsd.edu}
\affiliation{Department of Mechanical and Aerospace Engineering, University of California, San Diego, La Jolla, California 92093, USA}
\affiliation{Program in Materials Science and Engineering, University of California, San Diego, La Jolla, California 92093, USA}

\begin{abstract}
Spatio-temporal patterns emerging from an initial quiescent, uniform state is a phenomenon observed in many dynamical systems sustained far from thermodynamic equilibrium, the practical application of which has only recently begun to be explored.
As the underlying dynamics are typically complex, pattern formation is often theoretically analyzed and understood via phenomenological models, which effectively represent the causal mechanisms, but obscure the link between the small-scale interactions/processes and the observed macroscopic behavior.
Moreover, efforts to prescribe the patterning response are often undercut by the experimental inaccessibility of the small-scale constituents/processes.
This article demonstrates an artificial system (i.e., a robotic mechanical metamaterial) as an accessible and versatile platform within which to explore and prescribe the patterning response of non-equilibrium systems.
%Here, we construct an artificial, tunable non-equilibrium system (i.e., a robotic mechanical metamaterial) that numerically and experimentally exhibits a patterning response, which the corresponding first-principles model anticipates.
Specifically, in varying a feedback parameter within the prescribed reaction kinetics, the robotic mechanical metamaterial alternately develops spatial and temporal oscillations in the displacement field following a perturbation of the initial quiescent, uniform state.
%The results position robotic metamaterials as a versatile platform for studying and engineering non-equilibrium phenomena at the macroscale.
The platform is amenable to a first-principles analytical description so that corresponding theoretical results possess qualitative and quantitative significance, and maintain connection to the specific system parameters.
%Further development may 
\end{abstract}

\maketitle

% \red{Physical Review Applied - Letter - 3500 words. Scope - Nonlinear dynamics and pattern formation in natural or manufactured systems}

% \red{Nature Communications - Title - 15 words, Abstract - 150 words, Unreferenced main text - 5000 words, Figures+tables - Max. 10 display items, Captions - 350 words, Methods - 3000 words, References - 70 max.}

%\red{In certain scenarios (e.g., ), these patterns demonstrate some utility.}.
%Moreover, the response is not readily engineered be engineered.
%Introduce our system...

\section{Introduction}
Pattern formation\cite{TuringPTRSB1952,CrossRMP1993,GollubRMP1999} -- the autonomous emergence of structure from a uniform initial state -- is a phenomenon observed in many non-linear dynamical systems sustained far from thermodynamic equilibrium, spanning broad spatio-temporal scales and several scientific disciplines (e.g., chemistry\cite{ZhabotinskyJTB1973,PearsonScience1993,LeeNature1994}, cell biology\cite{HoferPD1995,ToriiTCB2012} and ecology\cite{GePNAS2023}, astrophysics\cite{SteinAJ1998}).
Beyond scientific curiosity, understanding pattern formation is of practical interest, e.g., for developing methods of diagnosing and treating cardiac arrhythmia\cite{WeiseThesis2012}, and predicting the distribution of biomass in water-scarce ecosystems\cite{ZelnikJTB2017}.
The engineering potential of stationary patterns has also been demonstrated, e.g., via the fabrication of nano-filtration membranes for water purification\cite{TanScience2018} and the design of surface textures for prescribed anisotropic deformation of inflatable structures\cite{TanakaSA2023,TanakaSR2024}.

Generally, patterns emerge due to an advection-/diffusion-driven instability, which begets spatially heterogeneous modes that peculiarities of the non-equilibrium evolution selectively reinforce.
%A classic example is that of a reaction that produces two chemical substances, one that stalls the reaction and diffuses rapidly,
%%(inhibitor),
%and a second that promotes the reaction and diffuses slowly,
%%(activator),
%%constituting a self-regulating process that may
%yielding a co-continuous mixture of activate and inhibited response.
However, although the mechanisms of pattern formation are broadly understood, the complex dynamics occurring across disparate spatio-temporal scales within an arbitrary system
%-- originating at the component (e.g., molecular) level --
often pose limitations to theoretical understanding and, thus, to practical development.
Typically, the theoretical analysis proceeds from analytical models derived phenomenologically rather than from physical laws, which obscures the link between small-scale interactions/processes and the observed macroscopic performance.
Moreover, the conditions for pattern formation are often difficult to satisfy and custom patterning-capable systems difficult to engineer.
%, or to manipulate \red{toward a more favorable set}.
%and the system parameters difficult to manipulate toward a custom response.
%To foster a more fundamental understanding -- which, ultimately, assists practical efforts -- a first-principles model stem
%Non-equilibrium systems whose small-scale effects are completely accountable and computationally tractable are, therefore, desirable.
%To this end, relatively simple systems -- especially those able to be prescribed at the component level -- are attractive.

Metamaterials are a class of materials comprising synthetic building blocks that, at the material level, engender unusual, often counter-intuitive behavior; drawing
%qualities that have drawn
the attention of diverse scientific communities (e.g., optics\cite{SoukoulisNP2011,UrbasJO2016}, acoustics\cite{LuMT2009,MaSA2016}, mechanics\cite{LeeAM2012,BertoldiNRM2017}), and
%generated
generating proposals for novel applications and devices\cite{KaraaslanOC2017,WangSMS2019,AdhikariJIMSS2021,ChenS2012,XinjingISJ2019,JiangAM2018,CaiNP2007,ChenAPL2007,StengerPRL2012}.
In addition, certain characteristics of natural phenomena have been reproduced in the metamaterial context, providing opportunities to experimentally explore those features under more readily accessible conditions\cite{FrazierAM2017,KhajehtourianEML2022,SkjaervoNRP2020}.
The bulk of documented metamaterial behavior reflects an energy-minimizing tendency;
%is the result of energy-minimizing processes;
however, the performance of robotic metamaterials -- a type of active material comprising energy-generating building blocks with embedded control loops -- is, inherently, of the non-equilibrium regime and, thus, presents a platform upon which to realize, study, and, ultimately, control associated phenomena.
%Thus, robotic metamaterials provide a tailorable platform upon which to reliably realize, study, and, ultimately, control non-equilibrium patterning.
%Extending the robotic metamaterial concept to the subject of pattern formation in non-equilibrium systems \red{allows pattern formation to be studied from physical principals and, by preserving the structure-response relationship, to be engineered for applications, e.g., autonomous morphable surfaces\cite{MortonJCM2023,JanbazNComm2024}}.

\begin{figure*}
    \centering
    \includegraphics{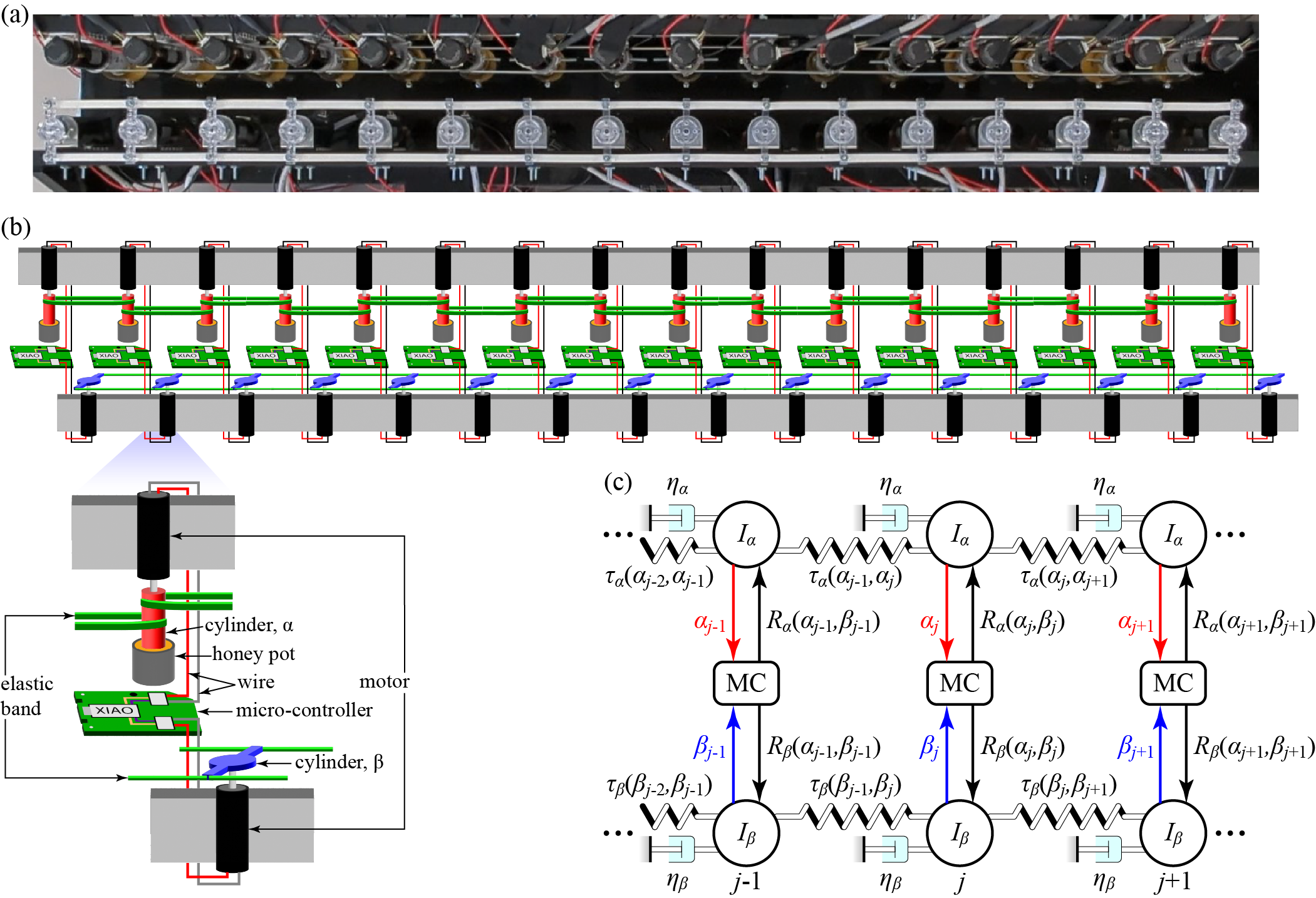}
    \caption{Robotic Metamaterial. (a) Overhead photograph of experimental set-up of robotic metamaterial test specimen comprising two arrays of elastically-coupled, motorized cylinders -- distinguished by rotations, $\alpha$ and $\beta$ -- with local inter-array interaction mediated by a micro-controller (MC).
    (b) Corresponding test specimen schematic with detail of unit cell assembly. (c) Lumped-parameter representation of the test specimen and Eq. \eqref{eq:Gov_Eqn_Disc}.
    }
    \label{fig:RMM_Schematic}
\end{figure*}

To date, robotic metamaterials have realized, e.g., the mechanical analogue of the non-Hermitian skin effect as a consequence of broadband non-reciprocal acoustic transmission\cite{BrandenbourgerNC2019}, adaptive locomotion via limit cycle deformations\cite{VeenstraNature2025}, and shape and strength modulation by mimicking cellular interactions in embryonic tissues\cite{DevlinScience2025}.
Here, we construct a robotic mechanical metamaterial from elastically-coupled active components (i.e., motors) that inherently maintain the system in a non-equilibrium state conducive to pattern formation as observed in chemical and biological systems.
The metamaterial experimental setup demonstrates a robust and predictable pattern forming ability in the displacement phase when subject to perturbations from the thermodynamic equilibrium state.
These experimental results are compared with analytical and numerical ones stemming from a physics-based model.

\section{Results}
\subsection{Pattern-forming Robotic Metamaterial}
Figure \ref{fig:RMM_Schematic}b displays a schematic of the robotic metamaterial test specimen (Fig. \ref{fig:RMM_Schematic}a), which comprises two interacting subsystems (i.e., discrete chains) of elastically-coupled active elements (i.e., 3D-printed cylinders affixed to the shafts of rotary motors) distinguished by the local rotational freedoms, $\alpha_j$ and $\beta_j$, respectively, where $j\in[\![1,16]\!]$ is the unit cell index.
Within each subsystem, the cylinders experience a non-linear coupling torque, $\tau_\blacksquare(\blacksquare_j,\blacksquare_{j+1})$, stemming from pre-stretched elastic bands of linearized rotational stiffness, $k_\blacksquare$, where $\blacksquare\equiv\alpha,\beta$.
In addition, the cylinders are subject to a local, non-conservative inter-chain reaction torque, $R_\blacksquare(\alpha_j,\beta_j)$, which is imposed by the attached motors under the direction of a programmable micro-controller that receives the current system state from the motor encoders.
The dynamics of an arbitrary unit cell are described by:
\begin{table}[b]
    \raggedright
    \caption{Robotic Metamaterial Physical Properties}
    \begin{tabularx}{\columnwidth}{>{\raggedright\arraybackslash}X >{\centering\arraybackslash}X >{\centering\arraybackslash}X >{\centering\arraybackslash}X}
        \toprule
        & $I$ ($\mathrm{kg\cdot m}^2$)& $\eta$ (N$\cdot$m$\cdot$s) & $k$ (N$\cdot$m)\\
        \midrule
        $\alpha$-chain & $6.39\times10^{-7}$ & $4.18\times10^{-4}$ & $1.46\times10^{-2}$\\
        $\beta$-chain & $3.13\times10^{-7}$ & $-$ & $4.57\times10^{-2}$\\
        \bottomrule
    \end{tabularx}
    \label{tab:Parameters}
\end{table}
\begin{subequations}
    \begin{align}
    I_\alpha\alpha_{j,tt}+\eta_\alpha\alpha_{j,t}=&-\tau_\alpha(\alpha_{j},\alpha_{j-1})-\tau_\alpha(\alpha_{j},\alpha_{j+1})\cdots\nonumber\\ 
    &+R_\alpha(\alpha_j,\beta_j),\\ 
    I_\beta\beta_{j,tt}+\eta_\beta \beta_{j,t}=&-\tau_\beta(\beta_{j},\beta_{j-1})-\tau_\beta(\beta_{j},\beta_{j+1})\cdots\nonumber\\
    &+R_\beta(\alpha_j,\beta_j),
\end{align}
\label{eq:Gov_Eqn_Disc}
\end{subequations}
\hspace{-5pt}where $I_\blacksquare$ and $\eta_\blacksquare$ are, respectively, the element inertia (i.e., cylinder and gearbox) and the local effective viscosity. Crucially, inertial effects are dwarfed by dissipative influences, rendering the dynamic response diffusive.
In addition, $R_\blacksquare(\alpha_j,\beta_j)$ is an (effectively) instantaneous reaction: compared to the micro-controller timing, the metamaterial response is protracted, a condition realized by submerging the cylinders of one subsystem -- here, chosen to be the $\alpha$-chain -- in a highly viscous liquid (i.e., honey).
Table \ref{tab:Parameters} collects the measured physical constants of the test specimen while $\tau_\blacksquare(\blacksquare_j,\blacksquare_{j+1})$ are functions defined by a polynomial fit to mechanical test data (see Supplemental Material).

For the purpose of analysis -- and consistent with strong elastic coupling, $k_\blacksquare$ -- Eq. \eqref{eq:Gov_Eqn_Disc} (\textit{sans} inertia) may be expanded in Taylor series to develop a corresponding continuum approximation.
For suitably smooth spatial variation in $\alpha_j$ and $\beta_j$, a 2nd-order expansion is sufficient, yielding:\setcounter{equation}{1}
\begin{subequations}
    \begin{align}
        \alpha_{,\bar{t}}&=\Delta \alpha+\bar{R}_\alpha(\alpha,\beta),\\
        \gamma \beta_{,\bar{t}}&=d\Delta \beta+\bar{R}_\beta(\alpha,\beta),
\end{align}
\label{eq:RD_Eqn}
\end{subequations}
%\hspace{-6pt}which affirms the negligible inertia and 
\hspace{-6pt}where $\bar{x}=x/a$, $\bar{t}=(k_\alpha/\eta_\alpha)t$, and $\bar{R}_\blacksquare(\alpha,\beta)$ are the dimensionless space, time, and inter-chain reaction functions; $\gamma=\eta_\beta/\eta_\alpha$ and $d=k_\beta/k_\alpha$ convey the relative temporal and diffusive characteristics of the two subsystems.
Written compactly, Eq. \eqref{eq:RD_Eqn} becomes $\mathbf{\Phi},_{\bar{t}}=\mathbf{D}\Delta\mathbf{\Phi}+\bar{\mathbf{R}}(\alpha,\beta)$ where $\mathbf{\Phi}^\mathrm{T}=[\alpha,\beta]$, $\mathbf{D}=\mathrm{diag}[1,d/\gamma]$, and $\bar{\mathbf{R}}^\mathrm{T}(\alpha,\beta)=[\bar{R}_\alpha(\alpha,\beta),\bar{R}_\beta(\alpha,\beta)/\gamma]$.
In the following, overbars are omitted to simplify the notation.

\begin{figure*}[t!]
    \centering
    \includegraphics{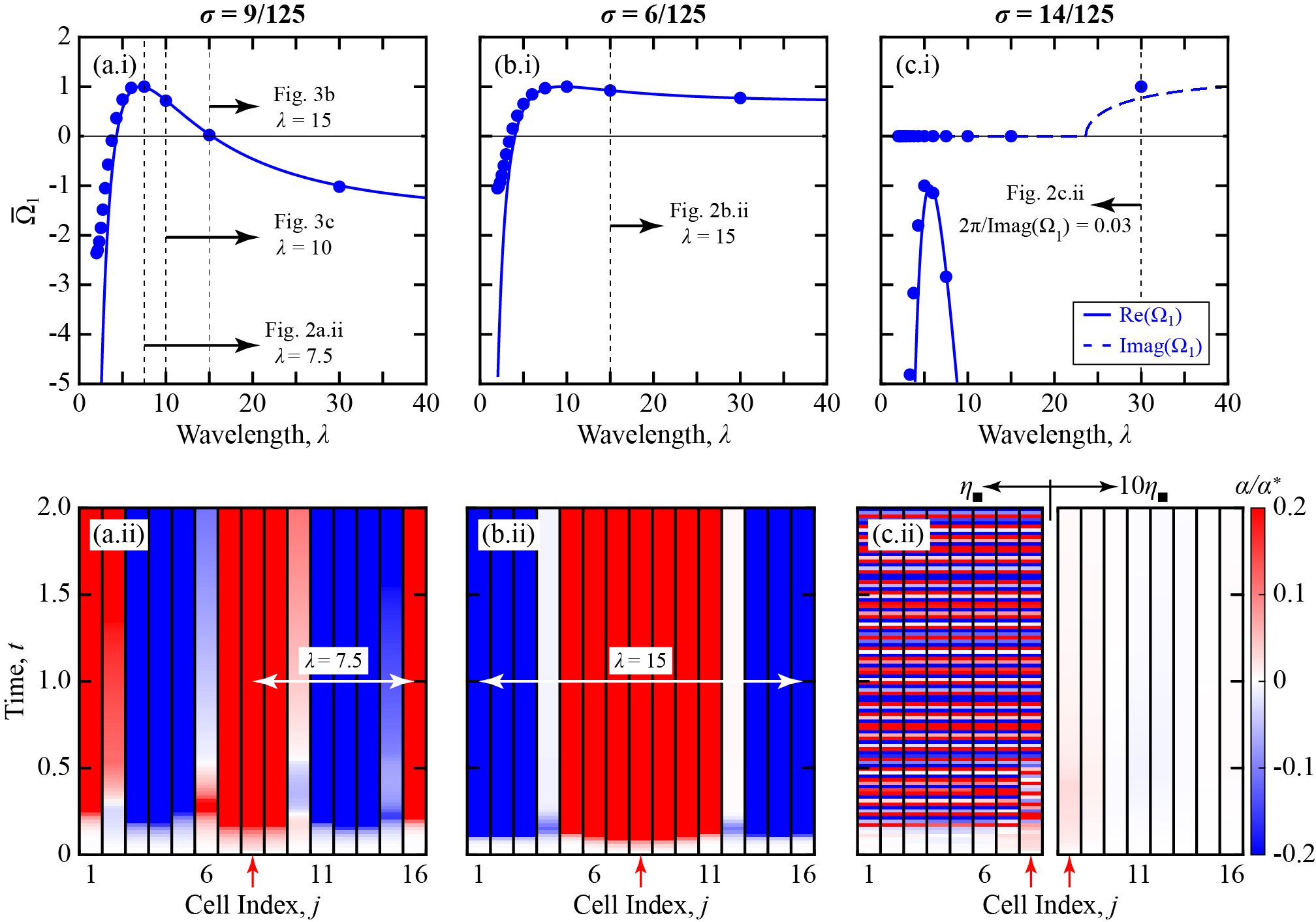}
    \caption{Effect of Feedback on Pattern Formation. (a.i--c.i) Results of linear stability analysis of the quiescent, spatially uniform solution, $\mathbf{\Phi}_0$, to Eqs. \eqref{eq:Gov_Eqn_Disc} (marker) and \eqref{eq:RD_Eqn} (curve) under different $\sigma$, representing the discrete and corresponding continuous systems, respectively. (a.i) With $\sigma=9/125$, the conditionally unstable system has a finite range where $\bar{\Omega}_1>0$; (b.i) with $\sigma=6/125$, the unconditionally unstable system has a semi-infinite range where $\bar{\Omega}_1>0$; and (c.i) with $\sigma=14/125$, the unconditionally stable system has $\bar{\Omega}_1<0$ over the whole domain. (a.ii--c.ii) Simulated response of the metamaterial test specimen to a perturbing torque applied to $\alpha_8$ (arrow). The wavelength of the spatial pattern that develops in (a.ii) and (b.ii) may be anticipated from the analysis in (a.i) and (b.i), respectively. The period of the temporal pattern in the left panel of (c.ii), $2\pi/\mathrm{Imag}({\Omega}_1)=0.03$s, may be anticipated from the analysis in (c.i). (a--c) Experimentally-consistent parameters used throughout.}
    \label{fig:TuringCondition}
\end{figure*}

\begin{figure*}[t!]
    \centering
    \includegraphics{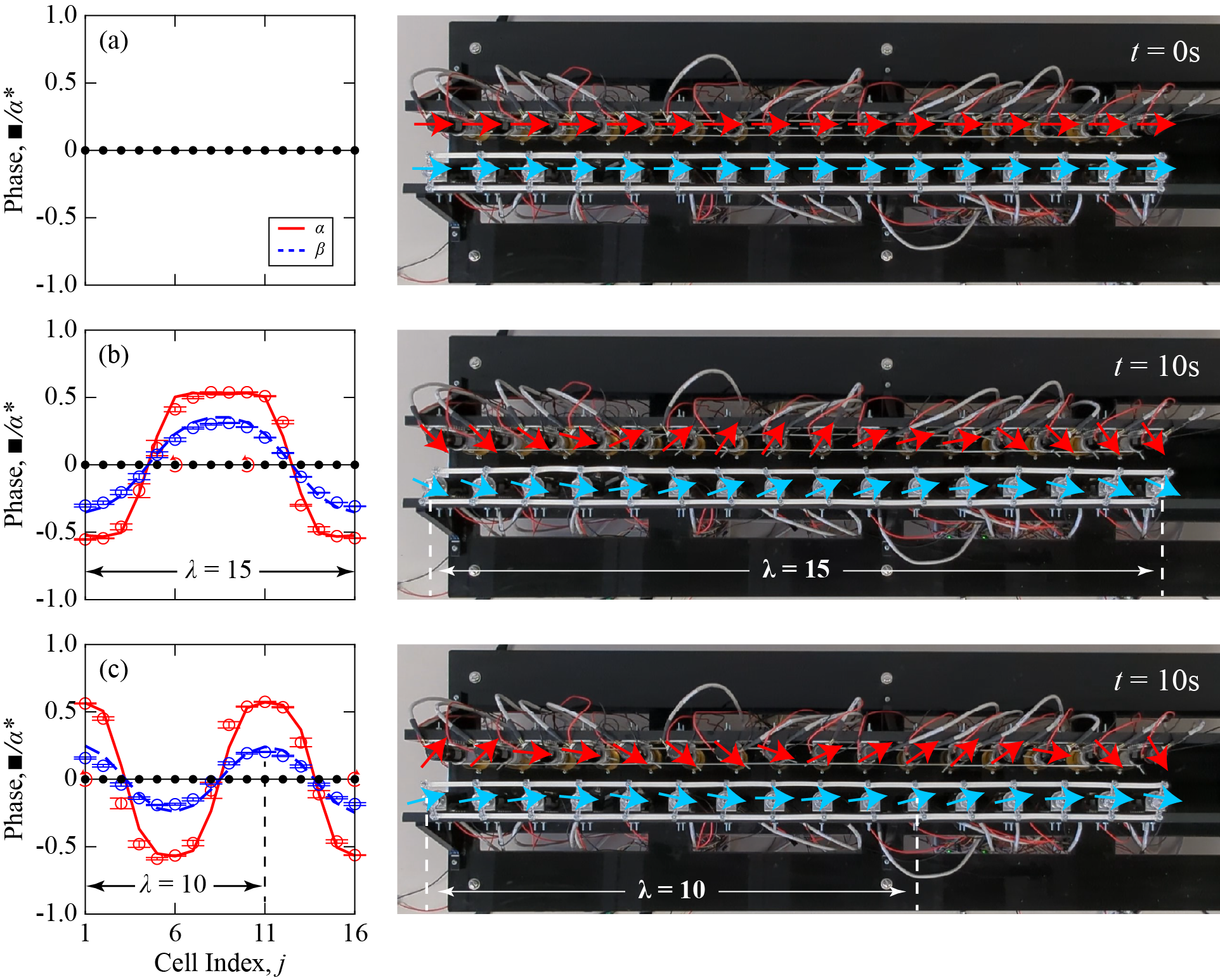}
    \caption{Pattern Formation in Robotic Mechanical Metamaterial. (a) The thermodynamic equilibrium state, $\mathbf{\Phi}_0:(\alpha_j,\beta_j)=(0,0)\;\forall\;{j}$, for the robotic metamaterial. Comparison of numerical (curve) and experimental (marker) cylinder angular displacements for (b) a CCW-CCW perturbation at sites $j=\{6,10\}$ (arrows), resulting in a stable spatial pattern of period, $\lambda=15$, (c) a CCW-CW perturbation at sites $j=\{1,16\}$ (arrows), resulting in a stable spatial pattern of period, $\lambda=10$. The experimental data is plotted as a mean phase with error bars indicating the range of response from multiple tests.}
    \label{fig:TuringPatterns}
\end{figure*}

\subsection{The Turing Mechanism}
Equation \eqref{eq:RD_Eqn} is of reaction-diffusion type\cite{LiPRE2013,TysonPhysD1988,ZhengCMA2015}, in particular, a model originally proposed by Alan Turing in 1952\cite{TuringPTRSB1952} to identify plausible mechanisms for the emergence of patterns in cellular differentiation during embryonic development (i.e., morphogenesis).
%Similar models facilitate the study of non-equilibrium dynamics in myriad diverse, complex systems.
In the present literature, models of this type -- which describe the non-equilibrium dynamics of myriad diverse, complex systems -- are often derived from phenomenological arguments rather than physical principles; however, by virtue of the engineered internal structure of the robotic metamaterial, each term in Eq. \eqref{eq:RD_Eqn} is readily traced to an underlying causal mechanism, thus preserving the structure-response relation and
%beyond a qualitative description,
promoting a quantitative understanding of the solution.
% -- attributes that may facilitate a custom dynamic response.

As observed by Turing, under the appropriate conditions, a small perturbation
%-- whose Fourier expansion is dominated by $\mathbf{\Phi}_\mathrm{p}=\tilde{\mathbf{\Phi}}\mathrm{e}^{\mathrm{i}\kappa{x}}$ --
from an initially quiescent, spatially uniform state, $\mathbf{\Phi}_0=\{(\alpha,\beta)|\mathbf{R}(\alpha,\beta)=\mathbf{0}\}$, will grow into a stable pattern in $\mathbf{\Phi}$.
%\cite{TuringPTRSB1952,KondoScience2010} in our mechanical system.
An examination of Eq. \eqref{eq:RD_Eqn} and the stability of the uniform solution, $\mathbf{\Phi}_0$, against perturbation reveals the necessary conditions for pattern development (see Supplemental Material):
\begin{subequations}
    \begin{align}
        \mathrm{tr}(\mathbf{J}_\mathbf{R}) &<0,\label{eq:TuringCond_a}\\
        \mathrm{det}(\mathbf{J}_\mathbf{R}) &>0,\label{eq:TuringCond_b}\\
        d R_{\alpha,\alpha}+R_{\beta,\beta} &> 2\gamma \sqrt{\mathrm{det}(\mathbf{D}) \mathrm{det}(\mathbf{J}_\mathbf{R})},\label{eq:TuringCond_c}
    \end{align}
    \label{eq:TuringCond}
\end{subequations}
\hspace{-3.5pt}where
%$\kappa=2\pi/\lambda$ is the spatial frequency and
$\mathbf{J}_\mathbf{R}$ is the Jacobian of $\mathbf{R}(\alpha,\beta)$ evaluated at $\mathbf{\Phi}_0$.
For conventional systems, conditions conducive to pattern formation might exist for only narrow ranges of the system parameters or within only isolated regions of the system domain, and may be difficult to sustain; however, the robotic metamaterial, by virtue of an accessible internal structure, does not inherit these challenges and is amenable to custom spatio-temporal response for applications (e.g., autonomous morphable surfaces).

En route to Eq. \eqref{eq:TuringCond}, the stability analysis yields the perturbation growth rates:\setcounter{equation}{3}
\begin{equation}
    \Omega_{1,2}(\kappa)=\frac{1}{2}\left(\mathrm{tr}(\mathbf{J}_\mathbf{\kappa})\pm\sqrt{\left[\mathrm{tr}(\mathbf{J}_\mathbf{\kappa})\right]^2-4\mathrm{det}(\mathbf{J}_\mathbf{\kappa})}\right),
    \label{eq:Omega}
\end{equation}
where $\kappa=2\pi/\lambda$ is the spatial frequency of the Fourier-expanded perturbation
%where $\kappa=2\pi/\lambda$ is the perturbation spatial frequency
and $\mathbf{J}_\mathbf{\kappa}=\mathbf{J}_\mathbf{R}-\kappa^2\mathbf{D}$.

Of the myriad possible definitions for the non-conservative reaction, $\mathbf{R}(\alpha,\beta)$, for demonstration, we choose to encode the following in the test specimen micro-controllers:
\begin{subequations}
    \begin{align*}
    R_{\alpha}(\alpha,\beta) & = k_{\alpha\beta} (\beta-\alpha)-\psi'(\alpha)-\sigma\beta,\\
    R_{\beta}(\alpha,\beta) & = k_{\alpha\beta} (\alpha-\beta),
\end{align*}
%\label{eq:ReactionTerms}
\end{subequations}
which, given $k_{\alpha\beta}>0$ and $\sigma>0$, classifies Eq. \eqref{eq:RD_Eqn}
as FitzHugh-Nagumo type\cite{FitzBMBP1955,FitzHughBJ1961,FitzMcGrawHill1969,NagumoPIRE1962}, a model for spiking neurons and the refractory period of cardiac cells.
Note that the first terms in $\mathbf{R}(\alpha,\beta)$ may also arise passively, e.g., by coupling each $(\alpha_j,\beta_j)$-pair in the test specimen via an elastic band of stiffness, $k_{\alpha\beta}$.
Similarly, there are several examples in the metamaterial literature\cite{FrazierAM2017,EichelbergPRAppl2022,RayPRM2025,CaoAFM2021} where $\psi'(\alpha)$, a non-monotonic function, is realized via mechanical and/or magnetic constituents.
Only the non-reciprocal term, $\sigma\beta$, which drives the system away from equilibrium, requires the integration of active components (i.e., motors) in what, otherwise, is a conventional energy-minimizing mechanical system.

Guided by Eq. \eqref{eq:TuringCond} and
%the already established
physical constants of the test specimen (Table \ref{tab:Parameters}), 
%we program the micro-controllers with parameters
we select the parameters for $\mathbf{R}(\alpha,\beta)$ that are conducive to the formation of spatial patterns, including $k_{\alpha\beta}=1/50$ and $\sigma=9/125$.
For simplicity, we encode $\psi'(\alpha)$ as an odd, piecewise continuous tri-linear function having three roots, $\alpha=\{0,\pm\alpha^*\}$.
In general, the slopes of the lines are given by:
\begin{equation*}
        \psi''(\alpha)=\begin{cases}
		k_2, & |\alpha|\leq k_1\alpha^*/(k_1-k_2),
		\cr k_1, & |\alpha|>k_1\alpha^*/(k_1-k_2),
	\end{cases}
\end{equation*}
where, in the present case, $\alpha^*=3\pi/8$ and $k_1=-k_2=3/50$.
%With the above dependencies and parameters,
Consequently, the initial uniform state of the test specimen is characterized by $(\alpha_j,\beta_j)=(0,0)\ \forall \ {j}$.
%and is conditionally stable.

Consider the effects of a disturbance to the uniform state; one that stimulates, e.g., a small region of non-zero $\alpha$.
According to Eq. \eqref{eq:RD_Eqn}, a positive (negative) $\alpha$ generates (diminishes) $\beta$ via the reciprocal interaction facilitated by $k_{\alpha\beta}$.
Meanwhile, for a positive (negative) $\beta$, the non-reciprocal term, $\sigma\beta$,
%does negative (positive) work on the system and
inhibits (promotes) the production of $\alpha$.
Given $\gamma\ll1$, fluctuations occur more rapidly within the $\alpha$-subsystem than within the $\beta$-subsystem.
Simultaneously, the specific non-linearity of $\psi'(\alpha)$ permits the existence of multiple stable solutions to Eq. \eqref{eq:RD_Eqn}: uniform and patterned.
%\blue{Simultaneously, the non-linear term, $\psi'(\alpha)$ promotes a self-driven, non-monotonic production (inhibition) of $\alpha$, allowing the existence of multiple stable, non-equilibrium global states -- uniform and patterned}.
%The non-linearity of $\psi(\alpha)$ provides for the existence of multiple stable solutions to Eq. \eqref{eq:RD_Eqn}, including the uniform and patterned states.
Given $d>1$, the diffusion of these effects occurs more rapidly within the $\alpha$-subsystem than within the $\beta$-subsystem.

%From the uniform state, consider the effects of a localized perturbation (e.g., a small region within which $\alpha\neq0$ manifests spontaneously).
%%From an examination of
%According to Eq. \eqref{eq:RD_Eqn},
%%a disturbance within, e.g., the $\alpha$-chain  generates/reduces $\beta$
%a positive (negative) $\alpha$ generates (reduces) $\beta$ via the reciprocal interaction facilitated by $k_{\alpha\beta}$.
%A positive $\beta$ extracts energy from the system via the non-reciprocal term, $\sigma\beta$, while the converse has the opposite effect.
%%Energy is extracted from (injected into) the system when $\beta$ is positive (negative) via the non-reciprocal term, $\sigma\beta$.
%Thus, a positive (negative) $\beta$ inhibits (promotes) the further production of $\alpha$.
%Given that $d>1$, a disturbance propagates slower within the $\alpha$-chain than within the $\beta$-chain.
%The non-linearity of $\psi(\alpha)$ provides for the existence of multiple stable solutions to Eq. \eqref{eq:RD_Eqn}, including the uniform and patterned states.

For select values of $\sigma$, Figs. \ref{fig:TuringCondition}a.i--c.i plot the max-normalized growth rate, $\bar{\Omega}_1=\Omega_1/|\max(\Omega_1)|$, of both the the discrete [Eq. \eqref{eq:Gov_Eqn_Disc}] and continuous [Eq. \eqref{eq:RD_Eqn}] systems, demonstrating excellent agreement that validates the approximation in Eq. \eqref{eq:RD_Eqn}.
In particular, for $\sigma=9/125$, Fig. \ref{fig:TuringCondition}a.i shows $\bar{\Omega}_1>0$ over a finite range, indicating $\mathbf{\Phi}_0$ to be conditionally stable and, therefore, receptive to only specific Fourier components of a broadband perturbation (i.e., pattern selection).
As a consequence of the accessibility of the internal architecture, compared to conventional non-equilibrium systems, this range is readily manipulated within the robotic metamaterial platform.
Conversely, Fig. \ref{fig:TuringCondition}b.i, where $\sigma=6/125$, depicts an extended plateau region in $\bar{\Omega}_1$ consistent with $\mathbf{\Phi}_0$ being generally unstable: a broadband perturbation will stimulate a broadband response.
For $\sigma=14/125$ [Fig. \ref{fig:TuringCondition}c.i] patterning is prohibited as $\mathrm{Re}[\bar{\Omega}_1(\kappa)]<0$ $\forall\kappa$, ensuring that the disturbed system ultimately returns to the initial quiescent, uniform state (i.e., $\mathbf{\Phi}_0$ is stable).
Moreover, beyond a critical wavelength, Fig. \ref{fig:TuringCondition}c.i shows $\bar{\Omega}_1$ acquires a non-zero imaginary component, signifying temporal oscillation in addition to decay.

\subsection{Simulation and Experiment}\label{sec:Simulation_and_Experiment}
The conditions for pattern development outlined in Eq. \eqref{eq:TuringCond} stem from a linear stability analysis of Eq. \eqref{eq:RD_Eqn}, which describes a continuous, hypothetically infinite system.
In the following, we validate these prescriptions in the context of the discrete, finite robotic metamaterial through experimental and numerical investigations.

From the quiescent, uniform state, short-duration perturbing torques are applied at specific sites within the test specimen (see MOV. 1) and the numerical model [Eq. \eqref{eq:Gov_Eqn_Disc}].
%\red{Within the numerical model [Eq. \eqref{eq:Gov_Eqn_Disc}], perturbations take the form of prescribed short-duration accelerations.}
The particular load distribution is chosen for the wavelength of its dominant Fourier component,
%based upon the wavelength of its anticipated dominant Fourier component,
which
%resides within the pattern grown region and
is predicted to characterize the periodicity of any emergent pattern.
For the test specimen, the torques are applied manually; therefore, in order to ensure the robustness of the metamaterial response against slight variations on the load distribution, tests at each loading scenario are repeated ten times.
Figure \ref{fig:TuringPatterns} compares the steady-state results, i.e., the $\mathbf{\Phi}_j$ reported by the motor encoders and that predicted by simulation.
In one scenario (Fig. \ref{fig:TuringPatterns}b), two counter-clockwise (CCW) torques of equal magnitude are applied at $\alpha_j$, $j=\{6,10\}$, a load distribution whose Fourier expansion is dominated by the symmetric component with $\lambda=15$.
In a second scenario (Fig. \ref{fig:TuringPatterns}c), the torques at $\alpha_j$, $j=\{1,16\}$ are of equal magnitude but opposite in sign (i.e., CCW-CW), stimulating the asymmetric Fourier component with $\lambda=10$.
In each case, as anticipated, the effect observed in experiment and simulation is a displacement pattern in $\mathbf{\Phi}_j$ with a spatial period corresponding to the loading.
Moreover, there is excellent agreement between the experimental and simulated results, which, therefore, connects the observed test specimen dynamics to theory [i.e., Eqs. \eqref{eq:TuringCond} and \eqref{eq:Omega}].

%thus $d\approx78/25$ and $\gamma\approx1/10$.

%\purple{To anticipate the response of the physical specimen, we simulate the evolution of its virtual counterpart as described by Eq. \eqref{eq:Gov_Eqn_Disc}.}

To complement the analytical results in Figs. \ref{fig:TuringCondition}a.i--c.i, we leverage the experimentally validated numerical model to simulate the evolution of the robotic metamaterial under different $\sigma$ following perturbation from the quiescent state.
For each scenario, we apply a short-duration perturbing torque at $\alpha_{8}$, which is associated with spatially-symmetric Fourier components with periods, $\lambda_n=15/n,\;n\in[\![1,8]\!]$.
In Fig. \ref{fig:TuringCondition}a.ii, a pattern characterized by $\lambda=7.5$ develops, representing the Fourier component with far and away the greatest growth rate.
In Fig. \ref{fig:TuringCondition}b.ii, the initial state is unstable such that the perturbation manifests a spatial pattern with $\lambda=15$ corresponding to the Fourier component with the
%most favorable combination of amplitude and growth rate.
greatest amplitude among competitors with similar growth rates.
Figure \ref{fig:TuringCondition}c.ii is split in order to depict the system response under two scenarios:
%with identical gain.
%highlight the competition between $I_\blacksquare$ and $\eta_\blacksquare$.
%highlight the consequences of
%In the left window,
in the left window, inertial and dissipative effects are comparable, leading to steady-state temporal oscillations (also observed in experiment); 
%In the right window,
in the right window, $\eta_\blacksquare$ are increased ten-fold while preserving $\gamma$ such that dissipative effects dominate, causing the perturbed system to decay back to $\mathbf{\Phi}_0$ in the steady state.

\section{DISCUSSION}
In closing, we propose a class of artificial systems characterized by an inherent far-from-equilibrium condition for the purpose of studying and engineering non-equilibrium phenomena at the macroscale.
%In closing, we propose a class of robotic mechanical metamaterials distinguished by a sustained far-from-equilibrium condition.
%as opposed to existing mechanical metamaterials  reliant upon energy-minimization\red{[cite]}.
%\red{conducive to the formation of spatial patterns.}
We realize a specific example in the form of a one-dimensional robotic metamaterial -- which, in principle, may be extended to high dimensions -- comprising two mechanical subsystems that interact non-reciprocally via embedded active elements.
As a result of the particular non-equilibrium architecture, following perturbation from an initial quiescent, uniform state, the assembled metamaterial test specimen and its virtual counterpart develop stable spatial and temporal patterns in the displacement field
%configuration parameters,
consistent with theoretical predictions.
This is significant as the first demonstration of a phenomenon commonly associated with, e.g., chemical and biological systems within a
%(artificial)
mechanical setting.
%which positions robotic metamaterials as a versatile platform for physically realizing and studying non-equilibrium phenomena at the macroscale.
Moreover, as the experimentally-verified analytical model proceeds from a first-principles approach,
%(rather than phenomenological),
%rather than a phenomenological one (the dominate paradigm),
it links the observed metamaterial response to its internal architecture.
%a rarity in studies of non-equilibrium media.
Such structure-response relations aid the development of robotic metamaterials for applications, e.g., autonomous morphable surfaces where patterns emerge as the cumulative effect of individual (linear) actuators simply responding to their local environment rather than coordinating their movements, an on-going robotics challenge.
Leveraging rapid developments in multi-scale manufacturing and stimuli-responsive materials, we envision future robotic metamaterials with smaller, more densely packed actuators powered by, e.g., pressure or temperature differences in the environment.

\section*{METHODS}

\subsection{Stability analysis}\label{subsec:Turing_AM}

The homogeneous equilibrium for the robotic metamaterial [Eq. \eqref{eq:RD_Eqn}] is calculated as $\mathbf{\Phi}_0:(\alpha_j,\beta_j)=(0,0)\ \forall \ {j}$ (see Supplemental Material). Linearizing the dynamics around the homogeneous equilibrium produces a diffusion matrix, $\mathbf{D}$ and reaction Jacobian matrix, $\mathbf{R}$ for the robotic metamaterial, analogous to those of the phenomenological reaction-diffusion model:
\begin{align*}
    \mathbf{D} = \begin{bmatrix}
    k_\alpha & 0\\
    0 & k_{\beta}/\gamma
    \end{bmatrix}, & \ \mathbf{R}=\begin{bmatrix}
    -k_{\alpha\beta}-\psi''(0) & k_{\alpha\beta}-\sigma\\
    k_{\alpha\beta}/\gamma & -k_{\alpha\beta}/\gamma
\end{bmatrix},
\end{align*}
where, $\gamma=\eta_\beta/\eta_\alpha<1$, and, $k_\alpha=\tau_{\alpha,\alpha}(0)$ and $k_\beta=\tau_{\beta,\beta}(0)$ denote the linearized torsional interaction stiffness around the equilibrium in the respective layers. Utilizing the above matrices in Eq. \eqref{eq:TuringCond}, we arrive at the conditions essential for conditional stability of the robotic metamaterial and choose the experimental parameters accordingly (see Supplemental Material).

\subsection{Mechanical components}
The active robotic metamaterial in Fig. \ref{fig:TuringPatterns} consists of two layers ($\alpha_j, \beta_j$) of 3D printed rotating cylinders coupled to neighboring sites via elastic bands and electrical circuits with a lattice parameter, $a=60 \mathrm{mm}$. Each layer consists of customized 3D printed cylinders (Fig. S1) with non-contact coupling between the layers via electrical circuit and motors. The coupling between adjacent unit cells is purely mechanical via elastic bands. In this section, we provide details of the mechanical and electrical setup used in the experiment.

Each lattice site in the $\alpha-\beta$ subsystems (Fig. S1), consists of (i) a 3D printed cylinder (Veroclear, Connex3 3D printer), (ii) a 3D printed custom mounting hub (Veroclear, Connex3 3D printer) connecting the cylinder and motor shaft, (iii) an aluminum bracket to fix the motor onto the vertical support (Pololu Item \#2676), (iv) a Brushed DC coreless motor (Assun Motors, AM-CL2242MAN 1205), (v) an optical encoder attached to the motor rear shaft (Anaheim Automation, ENC-A4TS-0100-079-M).

The $\alpha$-cylinders have a radius of $r=10 \mathrm{mm}$ and a height of $40 \mathrm{mm}$. The silicone bands in the $\alpha$-subsystem are secured onto the cylindrical surface via 1/2 inch, $\#4-40$ machine screws and hex-nuts. On the other hand, the $\beta$-cylinders are shorter in height, at $5 \mathrm{mm}$, with a circular center core, to be compatible with the mounting hubs and extended arms. The silicone bands in the $\beta$-subsystem are secured onto these extended arms via 1/2 inch, $\#4-40$ machine screws and hex-nuts at a radial distance, $2r$ to provide a higher effective torque (Fig. S3).

The silicone rubber bands used in both subsystems are hand cut from the Food Industry High-Temperature Silicone Rubber Sheet, Product \#86045K76 and \#86045K58 procured from  McMaster Carr. The elastic bands are cut to dimensions $83 \mathrm{mm} \times 6 \mathrm{mm} \times t$ and $55 \mathrm{mm} \times 6 \mathrm{mm} \times t$ for the $\alpha-\beta$ subsystems, respectively, where, $t=1/32$ inch. The length is chosen as, $L_\alpha=83 \mathrm{mm}$ and $L_\beta=55 \mathrm{mm}$ to ensure an initial pre-stretch around $10 \%$.

The complimentary pair of elastic bands between the interacting cylinders ensures that one of the elastic bands always remains in tension as the cylinders rotate, applying an equal and opposite elastic restoring torque on the connected cylinders. To increase the robustness of the numerical model and accurately capture the softening elastic behavior of the silicone bands, we perform Instron tests and evaluate the non-linear $3^\mathrm{rd}$ order polynomial approximations, $f_\alpha(\alpha)$ and $f_\beta(\beta)$ (Fig. S2) for the elastic response of the silicone bands. These functions are then utilized to calculate the elastic interaction torques, $\tau_\alpha(\alpha)$ and $\tau_\beta(\beta)$. Finally, to increase the discrepancy in the reaction-diffusion dynamics between the layers, the $\alpha$-cylinders are suspended over cups containing honey. The honey ensures increased viscous resistance to the $\alpha$-cylinder rotation, ensuring conditional stability of the thermodynamic equilibrium and the realization of rotational Turing patterns in the system. The viscous coefficient of honey is quantified by performing motor speed tests in the presence of honey (Fig. S6).

\subsection{Electrical components}

\emph{\bfseries Control System:} The control system for the inter-layer coupling between cylinders $\alpha_j$ and $\beta_j$ consists of (i) external power supplies (ii) an optical encoder attached to the motor rear shaft (ENC-A4TS-0100-079-M), (iii) a micro-controller (Seeeduino XIAO), (iv) two brushed DC motor drivers (Pololu DRV8874), and (v) two Brushed DC coreless motor (AM-CL2242MAN 1205).

\emph{\bfseries Power Supply:} We utilize two kinds of external power supplies - (1) a 5V-5A DC  (Pololu Item \#1462) power supply to power the encoders, micro-controllers and drivers and, (2) a 9V-30A DC power supply consisting of six 9V-5A DC (Pololu Item \#1465) power supplies connected in parallel to power the electric motors.

\emph{\bfseries Optical Encoders:} The optical encoder, ENC-A4TS-0100-079-M attached to the Brushed DC coreless motor, provides a precision of 400 counts per revolution (CPR), which translates to an angular resolution $0.0050\pi$ rad. An external 5V DC power supply powers all encoders. Each encoder outputs a pair of binary signals via the ENC A and ENC B terminals (see Fig. S7) that allow the respective unit cell micro-controller to process angular position of the attached motor.

\emph{\bfseries Micro-controller:} The Seeeduino XIAO micro-controller chip of a given unit cell, receives the encoder data from both the $\alpha-\beta$ subsystem electric motors and relays a PWM (pulse-width modulation) and direction signal to the motor drivers based on the feedback functions $R_\alpha$ and $R_\beta$. The micro-controller is also powered by the external 5V DC power supply. Even though the micro-controllers in each unit-cell don't need to be connected to an external computer to function, each micro-controller is connected to an $\mathrm{I}^2\mathrm{C}$ bus that relays the encoder data to a pair of central micro-controllers connected to a computer (See Fig. S7). This data bus enables the collection and graphical visualization of the steady-state encoder data from each lattice site after the system forms rotational Turing patterns.

\emph{\bfseries Motor Drivers:} The DRV8874 motor drivers, intake a SLP-PMODE input from the 5V DC power supply and a VIN-GND input from a 9V DC power supply. The SLP-PMODE input activates the motor drivers while the VIN-GND input is stepped down to the appropriate voltage level by the motor driver, based on the micro-controller PWM and direction signal input and supplied across the DC motor power terminals. The DRV8874 drivers are able to supply a maximum continuous current of $2.1$A at stall.

\emph{\bfseries Electric DC Motors:} The Brushed Coreless DC motors, AM-CL2242MAN 1205 are rated for a maximum stall torque, $\tau = 90$ mNm at $12$V and a stall current of $4.3$A. However, since the motor driver DRV8874 is rated for a $2.1$A continuous maximum current, we restrict the motor supply voltage at $5.86$V and stall torque at $44$ mNm via the micro-controller, to ensure a stall current below $2.1$A at all times and avoid any electrical damage.

\emph{\bfseries Software:} An Arduino software code is uploaded to each Seeeduino XIAO micro-controller to realize the feedback torque in both subsystems. The software code directs the micro-controller of a given unit cell, $j$ to receive the angular positions $\alpha_j$ and $\beta_j$, calculate the values $R_\alpha(\alpha_j, \beta_j)$ and $R_\beta(\alpha_j, \beta_j)$ and, relay the proportional levels of PWM and direction signals to the motor drivers and direct the drivers to supply the appropriate voltage across the motor terminals.

\subsection{Calibration and Measurements}
The elastic non-linear resistance, $f_\alpha$, and $f_\beta$ of the silicone bands are quantified by Instron tests. The detailed analysis and calculations of subsequent torque functions are elaborated in the Supplemental Material.

The Voltage, $\mathrm{V}_\mathrm{in}$ - no load speed, $\omega_\mathrm{NL}$ response of the brushed coreless DC motor is also tested at different applied voltages and verified with the parameters listed in the manufacturer data sheet. Similar tests are used to quantify the viscous coefficient of honey, $\eta_\alpha$ used a dissipative mechanism in the experiment. The $\mathrm{V}_\mathrm{in}$ - $\omega$ response in the presence of honey, is contrasted with the no-load motor data to quantify the viscous torque, $\mathbf{\tau}_\mathrm{visc}$ as a function of motor speed. The test data reveals a linear increase in $\mathbf{\tau}_\mathrm{visc}$ with $\omega$ allowing us to quantify the viscous effects of honey as the slope of this curve:
\begin{equation*}
    \eta_\alpha=\frac{\mathrm{d}\tau_\mathrm{visc}}{\mathrm{d}\omega},
\end{equation*}
A detailed analysis of the motor test data is available in the Supplemental Material.

\section*{Data availability}
The data that support the plots within this paper and other findings of this study are available from the corresponding author upon request.

\section*{Code availability}
The codes that support the plots within this paper, other findings of this study and the Arduino software code are available from the corresponding author upon request.

\section*{Acknowledgements}
M.J.F. acknowledges the support of the University of California start-up package and the Hellman Fellowship fund.
V.R. and M.J.F. thank Dr. Steve Roberts and Mr. Haoyi Tian for their skillful technical assistance in the experimental setup.
V.R. and M.J.F. thank Prof. N. Boechler for his insightful discussions.

\section*{Author contributions}
V.R. and M.J.F. developed the theoretical model and designed the experimental setup. V.R. conducted the experiments, and V.R. and M.J.F. performed the data analysis and wrote the paper.

\section*{Competing interests}
The authors declare no competing interests.

\section*{Additional information}
Supplementary information is available for this paper at \red{insert URL}.

\section*{Correspondence} 
Correspondence requests for materials should be addressed to M.J.F.

\section*{References}
% \bibliographystyle{ieeetr}
% \bibliography{References_MJF}

\newpage

\setcounter{equation}{0}
\renewcommand{\theequation}{S\arabic{equation}}
\renewcommand{\theHequation}{S\arabic{equation}}
\setcounter{figure}{0}
\renewcommand{\thefigure}{S\arabic{figure}}
\renewcommand{\theHfigure}{S\arabic{figure}}
\setcounter{section}{0}
\renewcommand{\thesection}{S\arabic{section}}
\renewcommand{\theHsection}{S\arabic{section}}
\setcounter{table}{0}
\renewcommand{\thetable}{S\arabic{table}}
\renewcommand{\theHtable}{S\arabic{table}}

\begin{widetext}
\noindent\textbf{\LARGE {Supplemental Information: Pattern Formation in Robotic Mechanical Metamaterial}}
\end{widetext}

\section{Robotic Metamaterial unit cell}

\begin{figure*}[t!]
\centering
\includegraphics{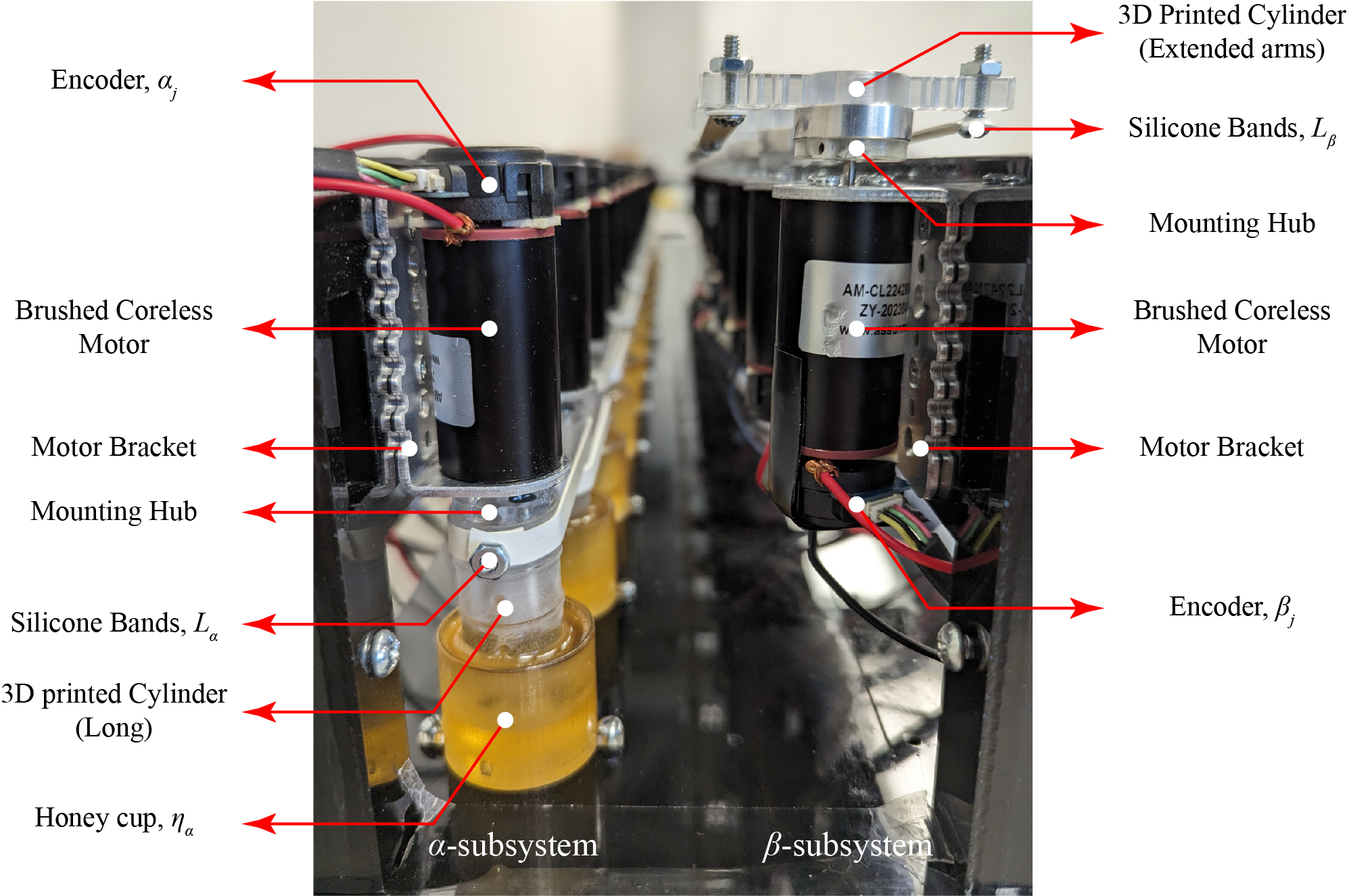}
\caption{Robotic Metamaterial components. Assembled view of an $\alpha-\beta$ unit cell labeling the electrical and mechanical components.}
\label{fig:SFIG1}
\end{figure*}

Fig. \ref{fig:SFIG1} shows an image of the assembled robotic metamaterial unit cell and labels the components constituting the lattice sites. Each subsystem consists of (i) a 3D printed cylinder (Veroclear, Connex3 3D printer), (ii) a 3D printed custom mounting hub (Veroclear, Connex3 3D printer) connecting the cylinder and motor shaft, (iii) an aluminum bracket to fix the motor onto the vertical support (Pololu Item \#2676), (iv) a Brushed DC coreless motor (Assun Motors, AM-CL2242MAN 1205), (v) an optical encoder attached to the motor rear shaft (Anaheim Automation, ENC-A4TS-0100-079-M).

The $\alpha$-cylinders have a radius, $r=10$ mm, and height, $\mathrm{H}=40$ mm. On the other hand, the $\beta$-cylinders have a radius of $8.5$ mm, height, $\mathrm{H}=5$ mm, and have two diametrically opposite extended arms with spring connection points at a radial distance of $2r$ from the center. The silicone bands in the $\beta$-subsystem are secured onto these extended arms at a radial distance, $2r$ to provide a higher effective torque.

The silicone rubber bands used in both subsystems are hand cut from Food Industry High-Temperature Silicone Rubber Sheet, Product \#86045K76 and \#86045K58 procured from McMaster Carr. The elastic bands are cut to dimensions $L_\alpha \times w \times t $ and $L_\beta \times w \times t $ for the $\alpha$ and $\beta$ subsystems, respectively, where, $w=6 \ \mathrm{mm}, \ t=3.125 \times 10^{-2} \ \mathrm{in}$ denotes the width and thickness of the silicone sheets. The lengths are chosen as $L_\alpha=83 \ \mathrm{mm}$ and $L_\beta=55 \  \mathrm{mm}$ to ensure an initial pre-stretch around $10 \%$ on both layers.

\section{Non-linear softening elastic behavior}

\begin{figure*}[t!]
\centering
\includegraphics{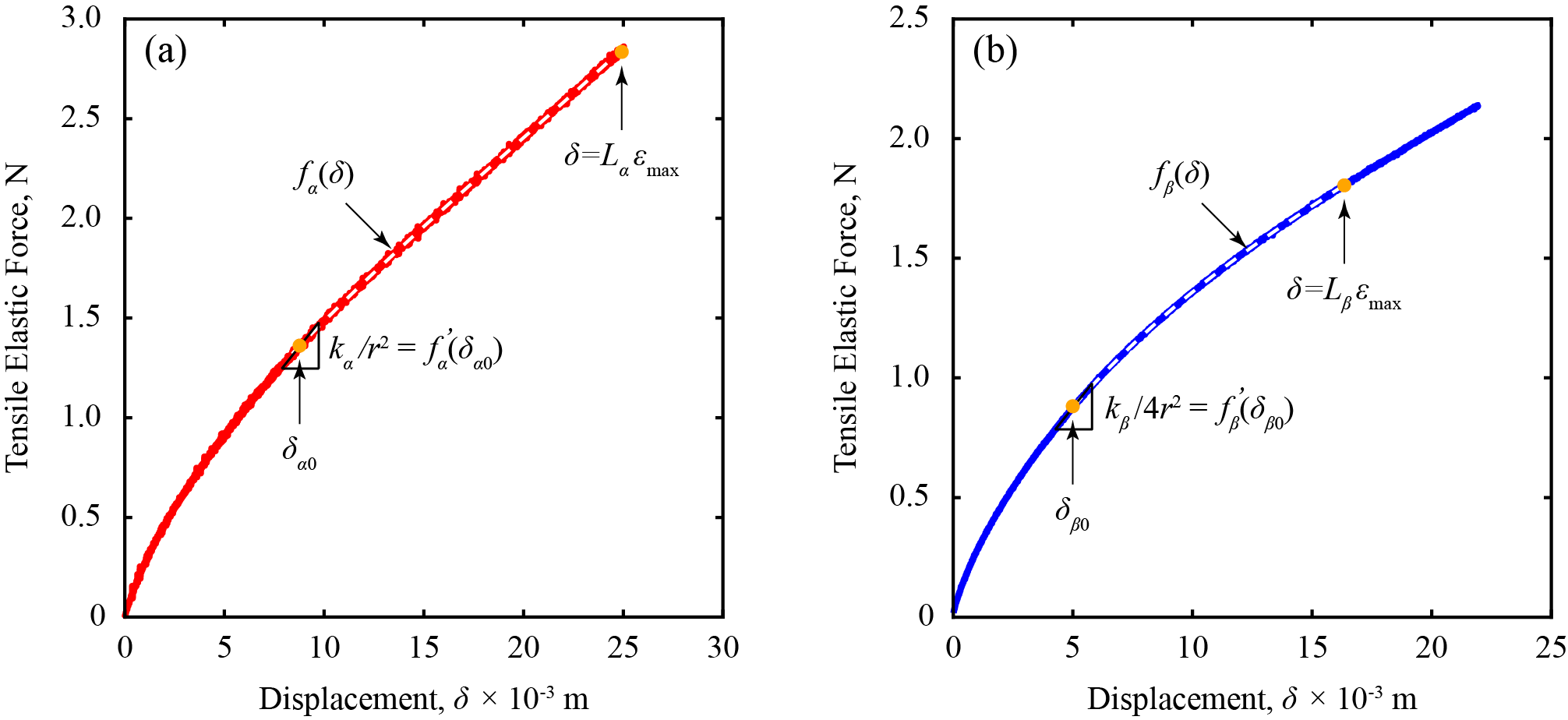}
\caption{Average Force-Displacement data from Instron tests for (a) $\alpha$-subsystem silicone bands and, (b) $\beta$-subsystem silicone bands.}
\label{fig:SFIG2}
\end{figure*}

Formulating the tensile resistance of the silicone bands as a linear spring force would fail to capture the softening material response over an extended deformation range. Hence, we perform tensile tests on multiple identical silicone bands using the Instron Universal Testing System to characterize their non-linear force-displacement behavior. This data is averaged for four tests and plotted for reference in Fig. \ref{fig:SFIG2}a,b, reinforcing a softening elastic response. The silicone bands are now integrated into our numerical model as non-linear springs whose force-displacement relation is modeled as a $3^\mathrm{rd}$ order polynomial approximations of the Instron data:
\begin{subequations}
\begin{equation}
    f_\alpha(\delta)=0.28+268.08\delta-8817.84\delta^2+123843.17\delta^3,
\end{equation}
\begin{equation}
    f_\beta(\delta)=0.19+164.96\delta-5965.25\delta^2+118186.51\delta^3,
\end{equation}
    \label{eq:ForceFunction}
\end{subequations}    
where $f_\alpha(\delta)$ and $f_\beta(\delta)$ denote the force functions in the respective subsystems. Fig. \ref{fig:SFIG2}a,b contrast $f_\alpha(\delta)$ and  $f_\beta(\delta)$ with the (averaged) raw Instron data for an experimental, accurately capturing the elastic response of these bands within a working strain range, $\varepsilon_0-\varepsilon_\mathrm{max}$. The lower limit,  $\varepsilon_0$ is taken as the pre-strain in the silicone bands in the initial configuration, and the upper limit, $\varepsilon_\mathrm{max}$ is chosen such that the actual strain experienced by the silicone bands in the experiment is well within this limit. 

In the $\alpha$-subsystem, the silicone bands wrap around a quarter sector length, $0.5r\pi$ of each connecting cylinder. Therefore, the strain limits are given as:
\begin{equation*}
    \varepsilon_0=\frac{\delta_{\alpha 0}}{L_\alpha}=\left(\frac{a+r\pi}{L_\alpha}-1\right) \approx 10 \%, \ \varepsilon_\mathrm{max} = 30 \%
\end{equation*}
where, $a=60$ mm denotes the lattice parameter, and $\delta_{\alpha 0}=a+r\pi-L_\alpha=8.4$ mm denotes the initial pre-stretch of the elastic bands. 

In the $\beta$-subsystem, the silicone bands attach to the extended arms of the cylinders, yielding a strain range:
\begin{equation*}
    \varepsilon_0=\frac{\delta_{\beta 0}}{L_\beta}=\left(\frac{a}{L_\beta}-1\right) \approx 9.1 \%, \ \varepsilon_\mathrm{max} = 30 \%
\end{equation*}
where, $\delta_{\beta 0}=a-L_\beta=5$ mm denotes the initial pre-stretch of the elastic bands.

%%%%% Figure command pasted here to position correctly in revtex reprint 2 column form

%%%%% Figure command pasted here to position correctly in revtex reprint 2 column form

\section{Elastic interaction torque}

Customized 3D-printed cylinders are attached to each motor shaft in each subsystem. The long cylinders of radius, $r$ (Fig. \ref{fig:SFIG1}) in the $\alpha$-subsystem enable them to comfortably sit in a cup filled with a viscous fluid, i.e., honey, and elastically interact with their intra-layer neighbors via the silicone bands. Alternately, the cylinders in the $\beta$-subsystem are short and contain extended arms with silicone bands connecting neighboring lattice sites at connection points at a radial distance, $2r$ (Fig. \ref{fig:SFIG1}) along these arms. This design is deliberately chosen for the following reasons:
\begin{itemize}
    \item Connections at larger radial distances from the motor shaft, translate to a higher torque.
    \item Higher torque generation allows us to use silicone bands with a lower initial pre-tension that minimizes potential bending effects, perpendicular to the motor shaft axis on these cylinders.
    \item The extended arms also serve as a visual representation of the rotational Turing patterns manifested by the robotic metamaterial.
\end{itemize}
\subsection{$\alpha$-subsystem}

\begin{figure}[t!]
    \centering
    \includegraphics{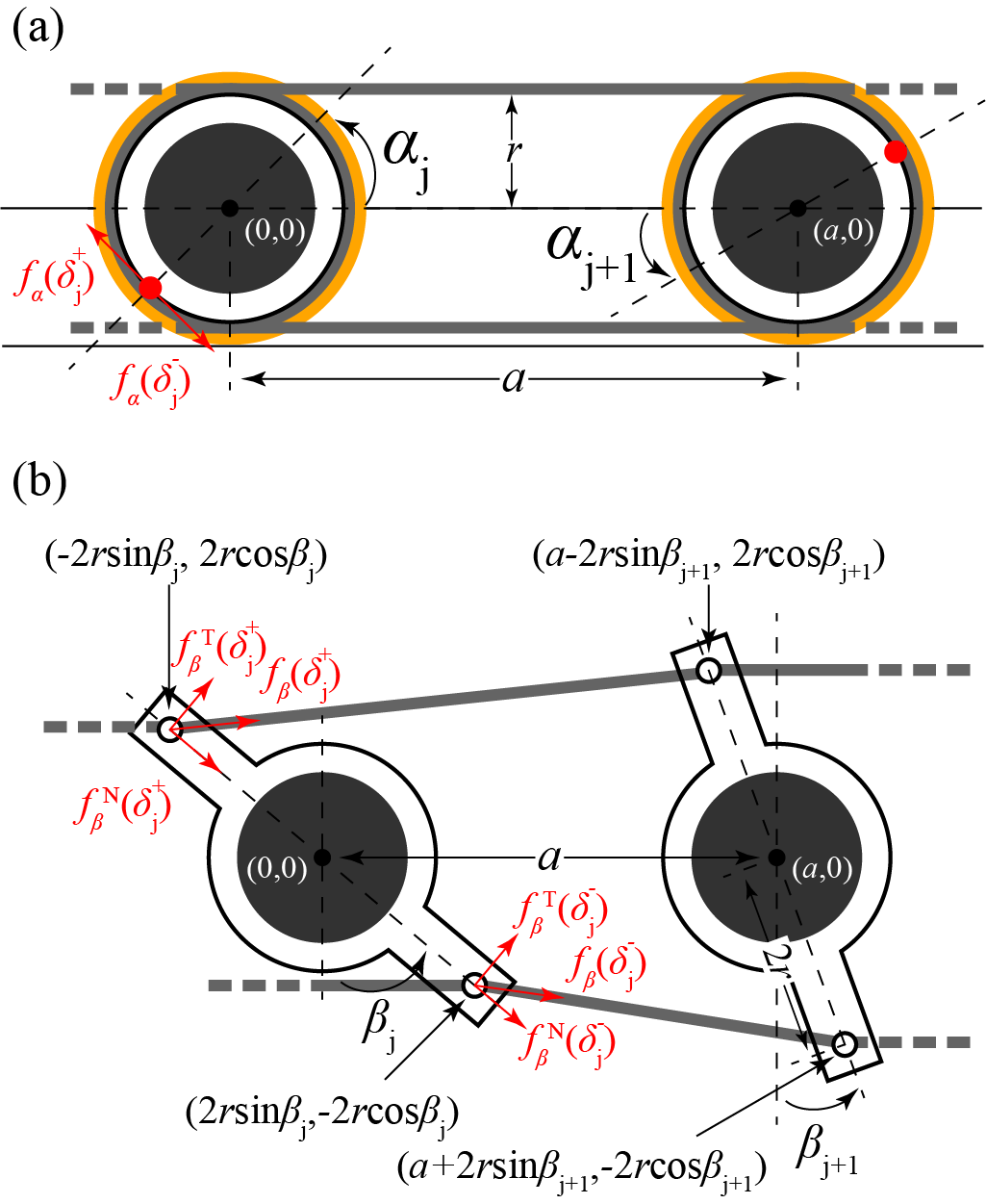}
    \caption{Interaction torque in the (a) $\alpha$-subsystem and, (b) $\beta$-subsystem.}
    \label{fig:SFIG3}
\end{figure}

Identifying the deformation of the pair of elastic bands between the $j^\mathrm{th}$ and $j+1^\mathrm{th}$ motors in the $\alpha$-subsystem as $\delta_j^\pm=\delta_{\alpha 0} \pm r(\alpha_j-\alpha_{j+1})$, the effective torque, $\mathbf{\tau_\alpha}$ applied on the $j^\mathrm{th}$ cylinder (Fig. \ref{fig:SFIG3}a) can be expressed as:

\begin{equation}
   \mathbf{\tau_\alpha}(\alpha_j,\alpha_{j+1})= \left\{
        \begin{array}{ll}
            \left[f_\alpha(\delta_j^+)-f_\alpha(\delta_j^-)\right]r, & \mathrm{if} \ \delta_j^\pm>0,\\[0.3cm]
             \hfill \left[f_\alpha(\delta_j^+)-f_\alpha(0)\right]r, & \mathrm{if} \ \delta_j^+>0, \delta_j^-<0,\\[0.3cm]
             \hfill \left[f_\alpha(0)-f_\alpha(\delta_j^-)\right]r, & \mathrm{if} \ \delta_j^+<0, \delta_j^->0,
        \end{array}
    \right.
    \label{eq:A_TorqueFunction}
\end{equation}

\subsection{$\beta$-subsystem}
A similar approach using $\delta_j^\pm=\delta_{\beta 0}\pm \Delta_\pm$ where, 
\begin{widetext}
\begin{equation*}
    \Delta_\pm=\sqrt{\left[a\pm2r(\sin\beta_j-\sin\beta_{j+1})\right]^2+\left[\mp 2r(\cos\beta_j-\cos\beta_{j+1})\right]^2}-a
\end{equation*}
can be used to establish the effective torque, $\mathbf{\tau_\beta}$ on the $j^\mathrm{th}$ cylinder (Fig. \ref{fig:SFIG3}b) in the $\beta$-subsystem:
\begin{equation}
   \mathbf{\tau_\beta}(\beta_j,\beta_{j+1})= \left\{
        \begin{array}{ll}
            \left[f_\beta^\mathrm{T}(\delta_j^+)-f_\beta^\mathrm{T}(\delta_j^-)\right]2r, & \mathrm{if} \ \delta_j^\pm>0,\\[0.3cm]
             \hfill \left[f_\beta^\mathrm{T}(\delta_j^+)-f_\beta^\mathrm{T}(0)\right]2r, & \mathrm{if} \ \delta_j^+>0, \delta_j^-<0,\\[0.3cm]
             \hfill \left[f_\beta^\mathrm{T}(0)-f_\beta^\mathrm{T}(\delta_j^-)\right]2r, & \mathrm{if} \ \delta_j^+<0, \delta_j^->0,
        \end{array}
    \right.
    \label{eq:I_TorqueFunction}
\end{equation}
where,
\begin{equation}
    f_\beta^\mathrm{T}(\delta_j^\pm)=f_\beta(\delta_j^\pm)\left(\frac{\cos\beta_j\left[a\pm2r(\sin\beta_j-\sin\beta_{j+1})\right] \mp 2r\sin\beta_j(\cos\beta_j-\cos\beta_{j+1})}{a+\Delta_\pm}\right),
\end{equation}
\end{widetext}
denotes the tangential component of the elastic force resulting in a torque on the $\beta$-subsystem.

\begin{figure}[h]
    \centering
    \includegraphics{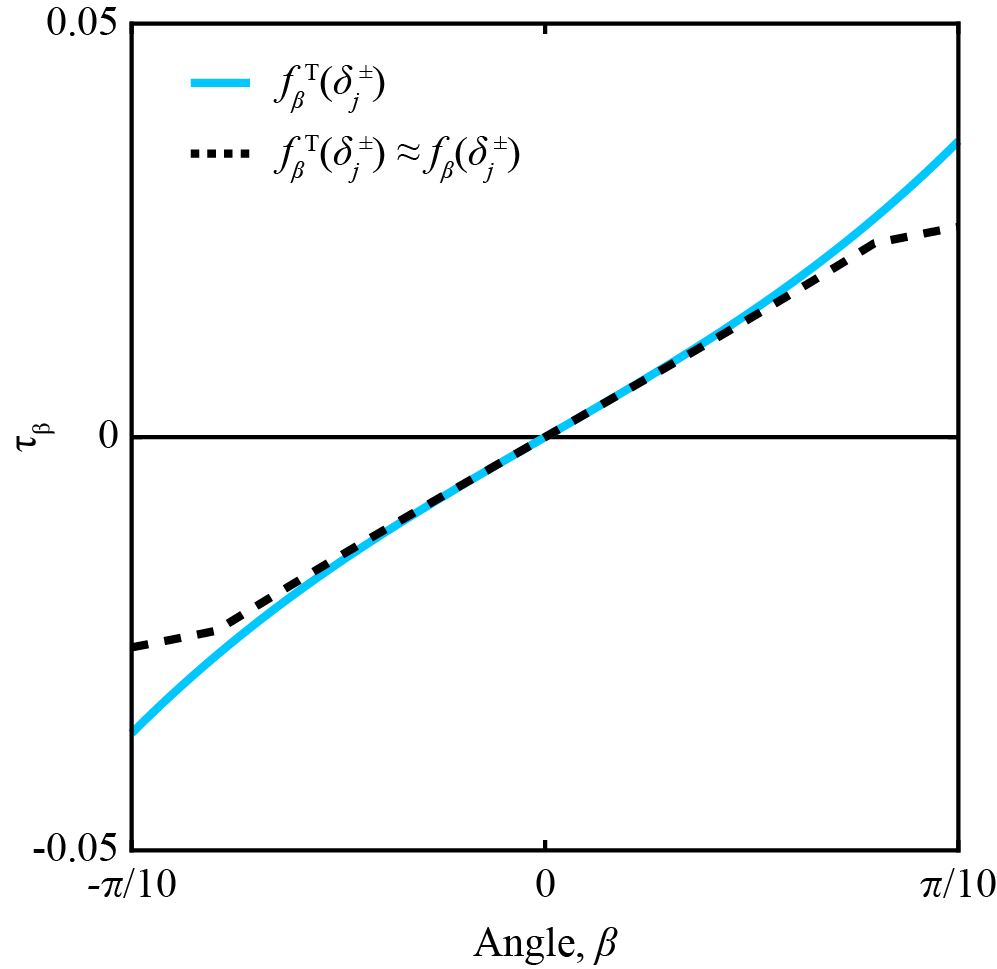}
    \caption{$\tau_\beta$ as denoted by Eq. \eqref{eq:A_TorqueFunction} and Eq. \eqref{eq:I_TorqueFunction}.}
    \label{fig:SFIG4}
\end{figure}
 If the change in rotations, $\beta_j-\beta_{j+1}$ is small, $\delta_j^\pm \approx \delta_{\beta 0}\pm 2r(\beta_j-\beta_{j+1})$ and $f_\beta^\mathrm{T}(\delta_j^\pm) \approx f_\beta(\delta_j^\pm)$, yielding the same form as Eq. \eqref{eq:A_TorqueFunction} for $\tau_\beta$. Fig. \ref{fig:SFIG4} contrasts the two forms of $\tau_\beta$, showing convergence between the forms for small $\beta_j-\beta_{j\pm1}$. As the $\beta$-subsystem is expected to have a lower rotational deformation than the $\alpha$-subsystem, we use the small angle form of $\tau_\beta$ in our numerical model to reduce the computation time. The excellent correlation between the experimental results and the numerical simulations in Fig. 3, in the main text reinforces the validity of this assumption.

\section{Electronic feedback functions}

\begin{align}
    R_\alpha(\alpha,\beta)&=k_{\alpha\beta}(\beta-\alpha)-\psi'(\alpha)-\sigma\beta\nonumber\\
    R_\beta(\alpha,\beta) & =k_{\alpha\beta}(\alpha-\beta),
    \label{eq:Feedback_Functions}
\end{align}

The active, non-reciprocal feedback functions, $R_\alpha$ and $R_\beta$ presented in Eq. \eqref{eq:Feedback_Functions}, are decisive parameters, governing the non-equilibrium response of the robotic metamaterial. The feedback function, $R_\alpha$ consists of three parts - a reciprocal interaction torque generated by a spring-like coupling of stiffness, $k_{\alpha\beta}$ between the $\alpha-\beta$ subsystems, a non-monotonic bistable torque, $\psi'(\alpha)$, and a linear feedback torque, $\sigma\beta$. The feedback function, $R_\beta$ contains only the reciprocal interaction torque. 

\begin{figure}[H]
    \centering
    \includegraphics{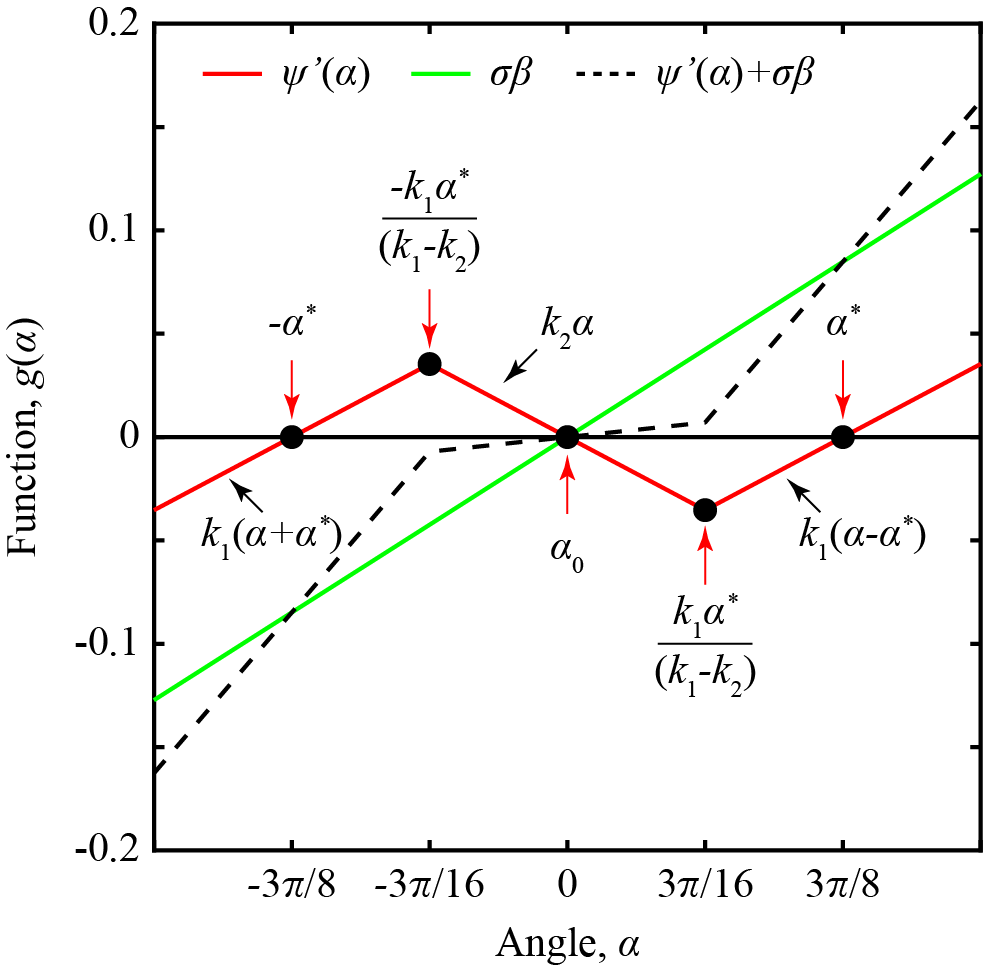}
    \caption{Electronic Feedback. Electro-mechanical equivalents to the reaction terms in phenomenological equations. Individual torque components, $\psi'(\alpha)$ and $\sigma\beta$ plotted for equilibrium criteria, $\beta=\alpha$, yielding a thermodynamic equilibrium, $\mathbf{\Phi}_0:(\alpha_j,\beta_j)=(\alpha_0,\beta_0)=(0,0)\ \forall \ {j}$ where, $\psi'(\alpha_0)+\sigma\beta_0=0$ (Eq. \eqref{eq:Equilibrium}).}
    \label{fig:SFIG5}
\end{figure}
Since the electronic setup operates on digital inputs and outputs, the electronic feedback can be readily correlated to a digital pulse-width modulation (PWM) value between $0-255$, provided the individual components of $R_\alpha$ and $R_\beta$ are linear functions of encoder readings, $\alpha$, and $\beta$. The reciprocal spring torque and the feedback are linear functions by design. So, to facilitate a one-to-one correlation between the expected motor torque and the PWM signal, we emulate the qualitative characteristics of a $\phi^4$-potential\cite{FeiPRE1993}, using a tri-linear torque function:

\begin{equation}
   \psi'(\alpha)= \left\{
        \begin{array}{ll}
            k_1(\alpha+\alpha^{*}), & \mathrm{if} \ \alpha < \frac{-k_1\alpha^{*}}{k_1-k_2},\\[0.3cm]
             \hfill k_2\alpha, & \mathrm{if} \ \frac{-k_1\alpha^{*}}{k_1-k_2} < \alpha < \frac{k_1\alpha^{*}}{k_1-k_2},\\[0.3cm]
             \hfill k_1(\alpha-\alpha^{*}), & \mathrm{if} \ \alpha > \frac{k_1\alpha^{*}}{k_1-k_2},
        \end{array}
    \right.
    \label{eq:TriLinear_Function}
\end{equation}
where, $\alpha=\pm\alpha^{*}$ and $\alpha=0$, respectively, represent the stable and unstable fixed points. The stable fixed points for the tri-linear function used in the experiment are located at $\alpha=\pm3\pi/8$. Fig. \ref{fig:SFIG5} plots the non-monotonic bistable torque and linear feedback functions used in the experiment.

%%%%% Figure command pasted here to position correctly in revtex reprint 2 column form
\begin{figure*}[t!]
\centering
\includegraphics{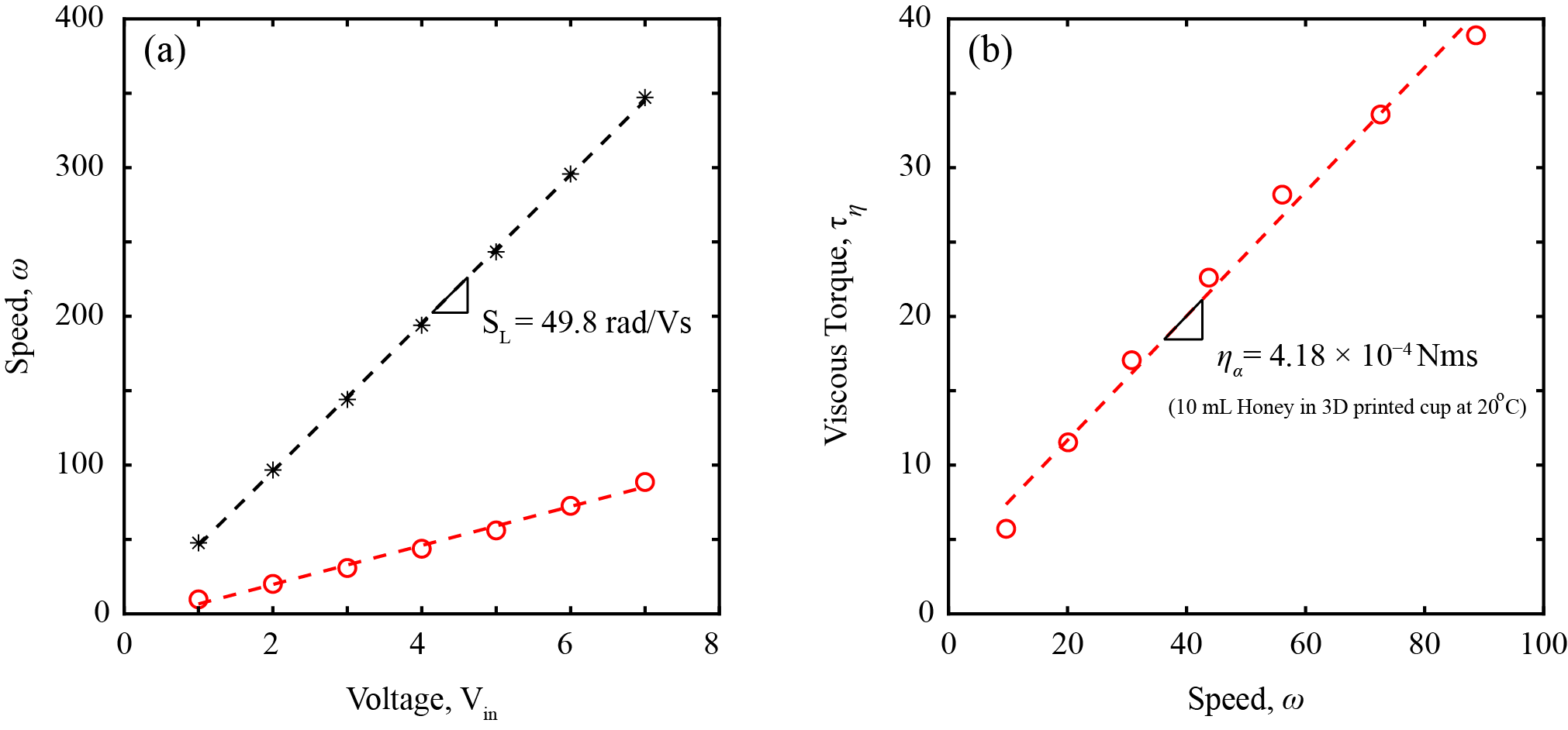}
\caption{Motor Characteristics and Viscous damping coefficient of Honey. (a) Voltage, $\mathrm{V}_\mathrm{in}$ - Speed, $\omega$ data with and without viscous effects of honey, (b) Motor speed, $\omega$ - viscous torque, $\tau_\eta$ of honey.}
\label{fig:SFIG6}
\end{figure*}

\section{Robotic Metamaterial - Conditional stability}

The thermodynamic equilibrium for the robotic metamaterial is calculated as:
\begin{align}
    R_\alpha(\alpha_0,\beta_0)&=k_{\alpha\beta}(\beta_0-\alpha_0)-\psi'(\alpha_0)-\sigma\beta_0=0,\nonumber\\[0.3cm]
    R_\beta(\alpha_0,\beta_0)&=k_{\alpha\beta}(\alpha_0-\beta_0)=0,
    \label{eq:Equilibrium}
\end{align}

Substituting Eq. \eqref{eq:TriLinear_Function} into Eq. \eqref{eq:Equilibrium}, we get $\mathbf{\Phi}_0:(\alpha_0,\beta_0)=(0,0)$. Performing a stability analysis of the thermodynamic equilibrium, $\mathbf{\Phi}_0$ by linearizing the interaction torque functions as linear spring torques with stiffness:

\begin{align}
    k_\alpha&=\frac{\partial \mathbf{\tau}_\alpha}{\partial \alpha}\bigg\vert_{\alpha=0}=r^2\frac{\partial f_\alpha}{\partial \delta}\bigg\vert_{\delta=\delta_{\alpha 0}}, \nonumber\\
    \ k_\beta&=\frac{\partial \mathbf{\tau}_\beta}{\partial \beta}\bigg\vert_{\beta=0}=4r^2\frac{\partial f_\beta}{\partial \delta}\bigg\vert_{\delta=\delta_{\beta 0}}\approx 3.12k_\alpha
\end{align}
for both subsystems (Fig. \ref{fig:SFIG2}a,b). Subsequently, the elastodynamic equation can be reformulated in the continuum form:

\begin{align}
        \eta_\alpha\alpha_{,t} & = k_\alpha\Delta \alpha+R_\alpha(\alpha,\beta)\nonumber\\
    \gamma \eta_\alpha \beta_{,t} & = k_\beta \Delta \beta+R_\beta(\alpha,\beta)
    \label{eq:Linear_EOM}
\end{align}
The choice of electronic motor feedback parameters, $k_{\alpha\beta}=1/50$, $\sigma=9/125$ and $k_1=-k_2=3/50$, interaction stiffness ratio, $k_\beta/k_\alpha=3.12$ and viscous coefficient ratio, $\gamma=\eta_\beta/\eta_\alpha\approx1/10$, facilitates the realization of a robotic metamaterial that satisfies the conditions necessary for conditional stability: 
\begin{subequations}
\begin{equation}   
    k_{\alpha\beta}\left(1+\frac{1}{\gamma}\right)+k_2>0,   
\end{equation}
\begin{equation}
     k_{\alpha\beta}(k_2+\sigma)>0, 
\end{equation}
\begin{equation}   
    -\left(\frac{k_\beta}{k_\alpha}\right)k_2-k_{\alpha\beta}\left(1+\frac{k_\beta}{k_\alpha}\right)>2\sqrt{k_{\alpha\beta}\left(\frac{k_\beta}{k_\alpha}\right) (k_2+\sigma)},    
\end{equation}
\label{eq:AM_TuringConditions}
\end{subequations}
Therefore, precisely perturbing the robotic metamaterial from its thermodynamic equilibrium manifests spatially periodic rotational \emph{Turing patterns} in the steady state.

%%%%% Figure command pasted here to position correctly in revtex reprint 2 column form
\begin{figure*}[t!]
    \centering
    \includegraphics{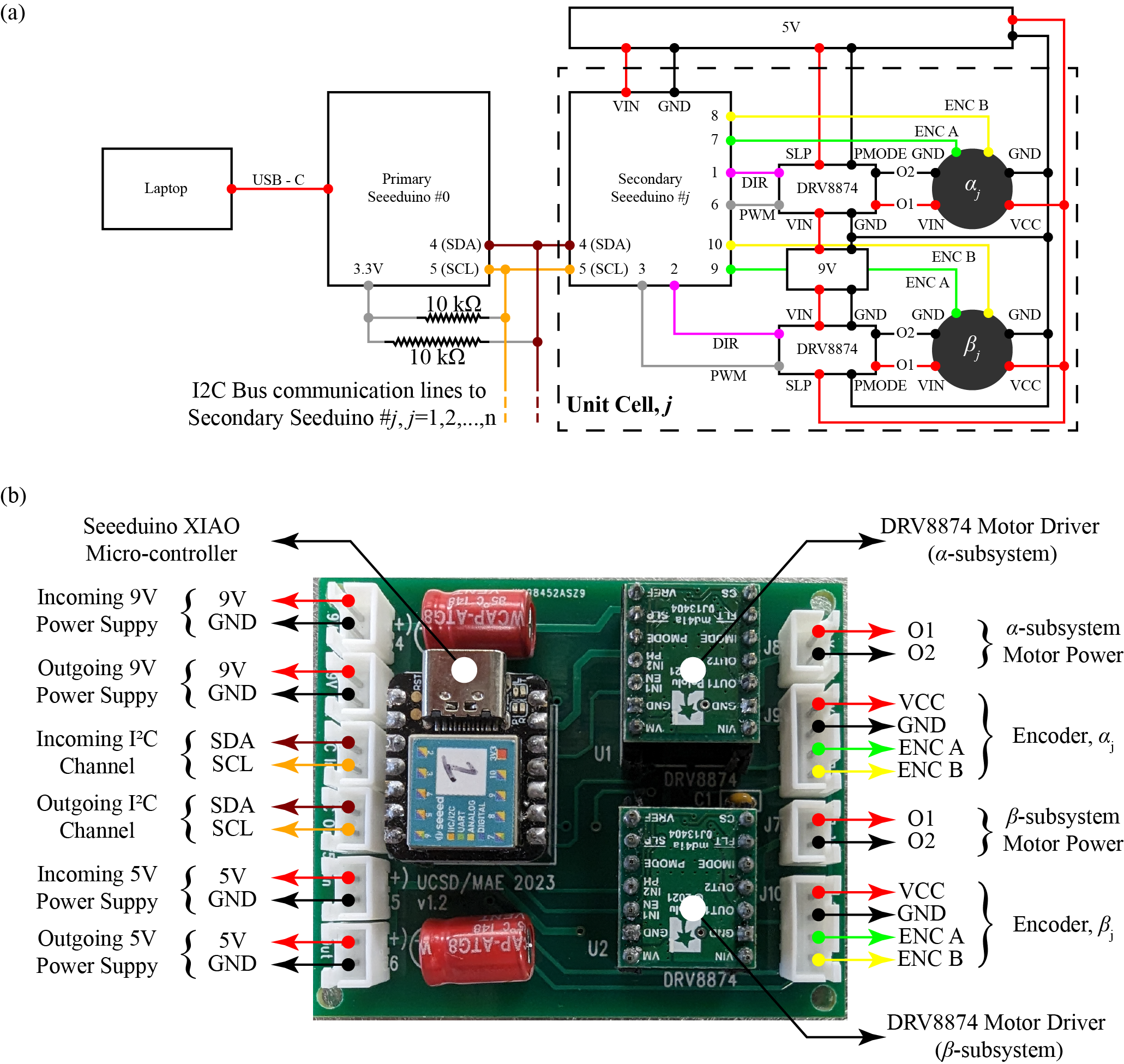}
    \caption{Electrical control system. (a) Electrical circuit diagram showing the internal connections within a unit cell, $j$ of the robotic metamaterial, (b) a custom-built PCB circuit board with the Seeeduino XIAO micro-controller and two DRV8874 motor drivers utilized as the unit cell control system in the experiment.}
    \label{fig:SFIG7}
\end{figure*}

\begin{figure*}[t!]
    \centering
    \includegraphics{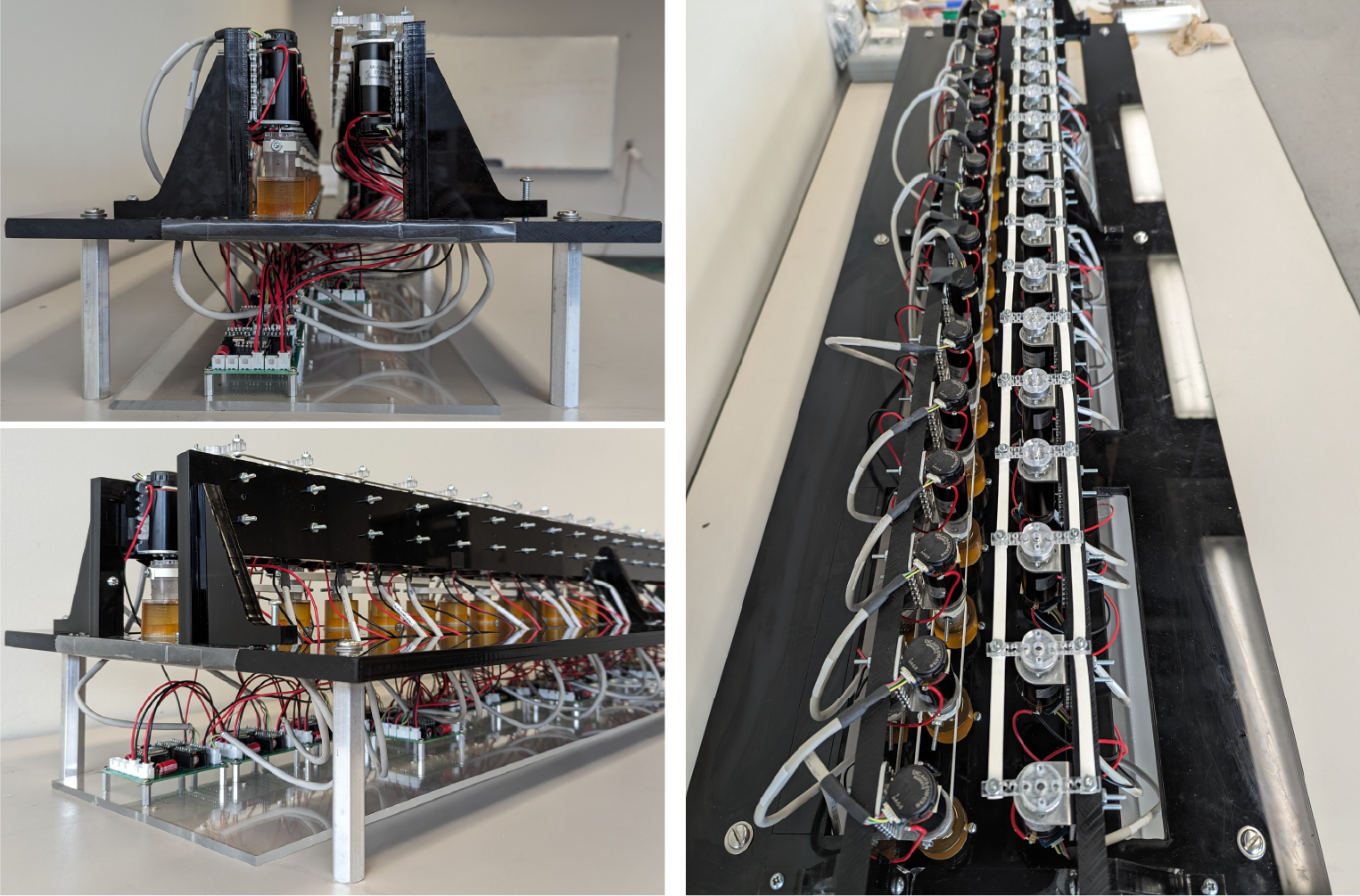}
    \caption{Snapshots of the experimental setup.}
    \label{fig:SFIG8}
\end{figure*}

\section{Motor function and Viscous behavior}

The brushed coreless motor speed is monitored at different voltages to characterize the input voltage, $\mathrm{V}_\mathrm{in}$ - no load speed, $\omega_\mathrm{NL}$ behavior, plotted as $\ast$-markers in Fig. \ref{fig:SFIG6}a. The experimental motor data suggests a linear increase in $\omega_\mathrm{NL}$ with applied voltage, $\mathrm{V}_\mathrm{in}$, generating a proportionality constant $\mathrm{S}_\mathrm{L}=49.8$ rad/Vs that is in good agreement with the manufacturer data, $\mathrm{S}_\mathrm{L}^*=47.9$ rad/Vs, verifying the rated motor characteristics. Therefore, using the calculated motor constant, $\mathrm{S}_\mathrm{L}$ and the motor voltage-stall torque rating of 12 V - 90 mNm, we formulate the typical linear output speed, $\omega$ - external load, $\tau_\mathrm{ext}$ relation for an applied voltage, $\mathrm{V}_\mathrm{in}$:

\begin{equation}
    \tau_\mathrm{ext}=\frac{0.09}{12}\left(\mathrm{V}_\mathrm{in}-\frac{\omega}{\mathrm{S}_\mathrm{L}}\right) \ \mathrm{Nm},
    \label{eq:Tau_ext}
\end{equation}

Note that, in the absence of any external torque or load on the motor shaft, i.e., $\tau_\mathrm{ext}=0$, we recover:

\begin{equation*}
    \omega=\omega_\mathrm{NL}=\mathrm{S}_\mathrm{L}\mathrm{V}_\mathrm{in},
\end{equation*}

The cylinders in the $\alpha$-subsystem experience an external viscous torque, $\tau_\eta$ as they are suspended in a honey cup. So, Eq. \eqref{eq:Tau_ext} will be utilized to determine the viscous coefficient of honey, $\eta_\alpha$. The motor output speed, $\omega$ with the attached cylinders immersed in honey is monitored at different applied voltages, $\mathrm{V}_\mathrm{in}$ and this data is plotted using o-markers in Fig. \ref{fig:SFIG6}a. The experimental data shows a visible decrease in output speed, $\omega$ compared to the no-load speed, $\omega_\mathrm{NL}$ at any given voltage, $\mathrm{V}_\mathrm{in}$, clearly indicating the effect of a dissipative viscous torque on the motor dynamics. The observed output speed, $\omega$ is substituted in Eq. \eqref{eq:Tau_ext} to determine the external viscous torque, $\tau_\eta$ experienced by the motor shaft and plotted as a function of the output speed, $\omega$ in Fig. \ref{fig:SFIG6}b. Surprisingly, the plot reveals a linear dependence of the viscous torque on the motor speed, despite honey being classified as a non-Newtonian fluid. We verify this correlation via multiple motor tests, allowing us to quantify the viscous coefficient of honey as the slope of this linear fit to the $\omega-\tau_\eta$ curve:

\begin{equation}
    \eta_\alpha=\frac{\mathrm{d}\tau_\eta}{\mathrm{d}\omega}=4.18 \times 10^{-4} \mathrm{Nms},
\end{equation}

The $\beta$-subsystem is not subject to any external viscous torque. Thus, the viscous coefficient in the $\beta$-subsystem is assumed to be an order of magnitude lower than the $\alpha$-subsystem, i.e., $\eta_\beta\approx0.1\eta_\alpha$, to ensure numerical stability of our time integration scheme\cite{NohCS2013}. 

\section{Electrical control system}

Fig. \ref{fig:SFIG7}a represents the electrical circuit diagram for the motor control system of an isolated unit cell that provides the active feedback torques $R_\alpha$ and $R_\beta$. Each unit cell of the robotic metamaterial comprises a secondary Seeeduino XIAO micro-controller, encoders, and motor drivers powered by a 5V-5A DC power supply and two brushed coreless motors ($\alpha_j$ and $\beta_j$) controlled via the respective motor drivers. The functioning of the control system is elucidated in the following iterative steps - 

\begin{enumerate}
    \item The Seeeduino micro-controller, $\#j$ receives the encoder positions, $\alpha_j$ (ENC A, ENC B - PIN $\#7$, PIN $\#8$) and $\beta_j$ (ENC A, ENC B - PIN $\#9$, PIN $\#10$) from the respective motors.
    \item The Seeeduino micro-controller, $\#j$ processes the active feedback torques, $R_\alpha$ and $R_\beta$ based on the encoder values.
    \item The Seeeduino micro-controller, $\#j$ then, relays a correlated direction (DIR) and pulse-width modulation (PWM) signal PIN $\#1,\#6$ and PIN $\#2,\#3$, respectively, to the DRV8874 motor drivers controlling the motors in the $\alpha-\beta$ subsystems.
    \item  Upon receiving the DIR and PWM signals from the micro-controller, the motor drivers step down the 9V input voltage and supply the appropriate voltage across the motor terminals (O1, O2 - VIN-GND).
    \item As the rotational phase of the robotic metamaterial evolves and the encoder readings change, the process repeats itself.
\end{enumerate}

The electrical circuit within a given unit cell, $j$ is self-sufficient and does not interface with the electrical circuits in the neighboring unit cells. However, to collect the encoder data from all unit cells, $j=1,2,...,n$ of the robotic metamaterial, the secondary micro-controller in each unit cell is connected to a primary Seeeduino micro-controller, $\#0$ via a 3.3 V, I$^2$C bus through the SDA-SCL pins (PIN $\#4$, PIN$\#5$) on each micro-controller. In addition, each secondary Seeeduino micro-controller is assigned a unique address, consistent with the unit cell number, $j=1,2,...,n$, to operate on this channel. The primary micro-controller then communicates with each secondary micro-controller in sequence to extract the encoder data of the associated unit cell and, relays the experimental data to a laptop via a USB-C connection, for graphical visualization and correlation with numerical simulations.

The internal connections between the micro-controllers, motor drivers, power supply lines, encoder cables, and motor power lines, in the electrical circuit in Fig. \ref{fig:SFIG7}a are custom printed on a PCB board shown in Fig. \ref{fig:SFIG7}b. The PCB also accommodates low capacitance capacitors across the 5V supply lines and high capacitance capacitors across the 9V supply lines and motor power lines, to protect the circuit from electrical damage due to any power surges during operation. The PCB design significantly simplifies the internal connections providing a reliable control system for the experiment.

\end{document}